\documentclass[nofootinbib,10pt,showpacs,tightenlines, showkeys, english,aps]{revtex4}
\usepackage[T1]{fontenc}
\usepackage[latin9]{inputenc}
\usepackage{color}
\usepackage{float}
\usepackage{epsfig,graphicx,times}
\usepackage{amstext}
\usepackage{amsmath}  
\usepackage{amssymb}
\usepackage{stackrel}
\usepackage{graphicx}
\usepackage{esint}
\usepackage{epstopdf}
\usepackage{epsfig}

\makeatletter
\@ifundefined{textcolor}{}
{%
 \definecolor{BLACK}{gray}{0}
 \definecolor{WHITE}{gray}{1}
 \definecolor{RED}{rgb}{1,0,0}
 \definecolor{GREEN}{rgb}{0,1,0}
 \definecolor{BLUE}{rgb}{0,0,1}
 \definecolor{CYAN}{cmyk}{1,0,0,0}
 \definecolor{MAGENTA}{cmyk}{0,1,0,0}
 \definecolor{YELLOW}{cmyk}{0,0,1,0}
}

\makeatother

\usepackage{babel}
\begin{document}

\title{Tomograms for open quantum systems: in(finite) dimensional optical
and spin systems}

\author{Kishore Thapliyal$^{a,}$\footnote{Email: tkishore36@yahoo.com}, Subhashish
Banerjee$^{b,}$\footnote{Email: subhashish@iitj.ac.in}, Anirban Pathak$^{a,}$\footnote{Email: anirban.pathak@gmail.com, Phone: +91 9717066494}}

\affiliation{$^{a}$Jaypee Institute of Information Technology, A-10, Sector-62,
Noida, UP-201307, India\\
 $^{b}$Indian Institute of Technology Jodhpur, Jodhpur 342011, India}
\begin{abstract}
Tomograms are obtained as probability distributions and are used to reconstruct a quantum
state from experimentally measured values. 
We study the evolution of tomograms for different quantum systems, both finite and 
infinite dimensional. 
In realistic experimental conditions, quantum states are exposed to the ambient environment
and hence subject to effects like decoherence and dissipation, which are dealt
with here, consistently, using the formalism of open quantum systems. This is extremely relevant from the perspective of 
experimental implementation and  issues related to state reconstruction in quantum computation and communication. These
considerations are also expected to affect the quasiprobability distribution obtained from experimentally generated tomograms 
and nonclassicality observed from them.
\end{abstract}



\maketitle

\section{{\normalsize{}Introduction}}

A quantum state can be characterized by a number of probability and
quasiprobability distribution functions \cite{st-est}. The quasiprobability
distributions are not true probability distributions as most of them
can have nonpositive values. Interestingly, this nonpositivity can be viewed as a signature
of nonclassicality or quantumness. Specifically, the negative values
of the Wigner \cite{wig} and $P$ \cite{glaub,sudar} functions serve as  
witnesses of nonclassicality. Further, zeros of $Q$ function \cite{comment1}  also  serve as 
 witness of nonclassicality. As there does not exist any 
straight forward prescription for direct measurement of these quasiprobability distributions, several efforts have been made to construct measurable probability 
distributions that can be used to uniquely construct either all or some of these quasiprobability distributions. Such measurable probability distributions 
are referred to as  tomograms  \cite{tom-review,with-adam,manko-fss,manko-spin-tom}. 
In other words, the tomogram is a scheme for measuring a quantum state by using a representation in one to one correspondence with the true probability
distribution rather than with a quasidistribution \cite{comment13-ref1}. 
 A relationship between a tomogram and a quasidistribution function, such as the
Wigner function, can be established for both continuous and discrete systems \cite{op-tom,fss-tom}. 
Specifically, in Ref. \cite{op-tom} it was shown that quasiprobability distributions ($P$, $Q$, and Wigner functions) can be uniquely determined in terms of 
probability distributions for the rotated quadrature phase which can be viewed as an optical tomogram of the state. Similarly, in Ref.  \cite{fss-tom} it was shown that for finite dimensional phase states, discrete Wigner functions and
tomograms are connected by a discretization of the continuous variable
Radon transformation and was referred to as the \textit{Plato transformation}.

In the recent past, a few successful attempts have been made to measure
Wigner function directly in experiments \cite{wig-exp,wig-exp2},
but the methods adopted are state specific. The same limitation is also valid for the theoretical
proposals  \cite{wig-diff} for the measurement of Wigner function. Further, optical homodyne tomography has been employed for the experimental measurement of the 
Wigner functions of vacuum and squeezed states in \cite{comment5exp-ref1,comment5exp-ref2},  while distributions corresponding to Pegg-Barnett and 
Susskind-Glogower phase operators were also obtained in \cite{comment5exp-ref2}. An experimental measurement of 
the $P$, $Q$ and Wigner quantum phase distributions for the
squeezed vacuum state has been reported in \cite{comment5exp-ref3}. Precision of homodyne tomography technique was compared with conventional detection techniques in 
\cite{comment5theory-ref1}. A number of alternative methods of tomography have also been proposed \cite{comment5theory-ref2,comment5theory-ref4,comment5theory-ref5}, and 
exploited to obtain phase distributions like Wigner and $Q$ functions \cite{comment5theory-ref3}. In \cite{CV-rev} continuous variable quantum state tomography was
reviewed from the perspective of quantum information.   
In brief, there does not exist any general prescription for direct
experimental measurement of the  Wigner function and other quasidistribution functions.  In practice, to detect
the nonclassicality in a system the Wigner function is obtained either
by photon counting or from experimentally measured tomograms \cite{wig-diff}. Thus, tomograms are very important for the
identification of nonclassical character of a physical system.
In another line of studies, simulation of quantum systems have been performed using tomography. For example, tomograms were used 
for simulation of tunneling \cite{comment9-ref1,comment9-ref2,comment9-ref4} and multimode quantum states \cite{comment9-ref3}. Attempts have also been made 
to understand the tomogram via path integrals \cite{comment9-ref5,comment9-ref6}. 

Furthermore, how to reconstruct a quantum state from experimentally measured
values is of prime interest for both quantum computation \cite{wig-exp}
and communication \cite{tel-143}. Specifically, in Ref. \cite{wig-exp} it is strongly established that 
tomography and spectroscopy can be interpreted as dual forms of quantum computation, and in Ref. \cite{tel-143}, quantum teleportation was 
experimentally performed over a distance of 143 km and the quality of teleportation was verified with the help of quantum process tomography (QPT) of 
quantum teleportation without feed-forward. Here it would be apt to note that QPT is an aspect of quantum state tomography in which a quantum process is obtained
as a trace preserving positive linear map \cite{QPT}.   In the recent
past, quantum process tomography has been discussed from the perspective of open
quantum system effects \cite{QPT-open,marzolino1,marzolino2}. 
A novel method of complete experimental characterization of quantum optical processes was introduced in \cite{comment8-ref1}. It was further developed 
in \cite{comment8-ref2, comment8-ref3} and extended to characterization of $N$-modes in \cite{comment8-ref7}. In \cite{comment8-ref4}, QPT was applied to 
the characterization of optical memory based on electromagnetically induced transparency while \cite{comment8-ref5} and \cite{comment8-ref6} were devoted to QPT
of the electromagnetic field and conditional state engineering, respectively.  
Quantum state tomography has its applications in quantum cryptography as well
\cite{tom-cryp}.  Specifically, in Ref. \cite{tom-cryp} an interesting  protocol of quantum cryptography was proposed  in which eavesdropping in the quantum channel was checked by requiring consistency of  outcome of the tomography  with the unbiased noise situation.   Keeping these facts in mind, we aim to construct tomograms
for a number of physical systems of practical relevance (mostly having applications in quantum computation and communication)
and investigate the effects of various types of noise on them.

From the experimental perspective, a quantum state always interacts
with its surroundings. Hence, the evolution of the corresponding tomogram
after taking into account the interaction of the quantum state with
its environment should be considered. This can be achieved with the
open quantum system formalism \cite{louis,bp,pi,sbqbm}. Specifically,
both purely dephasing (QND) \cite{QND} and dissipative \cite{SGAD}
open quantum system effects have been studied here. Interestingly,
both these effects have also been experimentally realized in the recent
past \cite{haroche,turchette}. In Ref. \cite{our-qd-paper}, a systematic
study of quasidistribution functions was made for a host of interesting
states under general open system evolutions. 

Here, we set ourselves the task of obtaining the tomograms for various
finite and infinite dimensional quantum systems in different open
quantum system scenarios. For finite dimensional spin states,
the tomogram is the distribution function of the projections of the
spin on an arbitrary axis, characterized by Euler angles, and can
be obtained from the diagonal elements of the rotated density matrix, while for continuous variable systems, such as the radiation field, the 
analog would be the homodyne probability. It follows from general group theoretical arguments that, 
making use of unitary irreducible square integrable representation of the tomographic group under consideration, a unified tomographic
prescription can be developed for both finite dimensional and continuous variable systems \cite{spin-ariano}.  
Tomograms for spin states have been developed both as projections on an arbitrary axis \cite{comment6-ref1} as 
well as by using a discrete variable analog of symplectic 
tomography \cite{comment6-ref2}.  
 Tomograms of optical systems have
been well studied in the past \cite{op-tom,dodonov-manko,tom-review,CV-rev,paris-rev}.
In Ref. \cite{mar-non-mar} quantum state tomography was used to determine the degree
of non-Markovianity in an open system. Further,  thermal noise is used in tomography (for reconstruction of photon number 
distributions) as a probe \cite{tom-by-noise}.

The paper is organized as follows. In Section \ref{sec:Tomograms-of-single-half},
tomograms of single spin-$\frac{1}{2}$ (qubit) atomic coherent state
under purely dephasing (QND) and dissipative evolution are obtained.
Further, the tomogram of two spin-$\frac{1}{2}$ (qubit) quantum state
is studied in Section \ref{sec:Tomogram-of-two-spin} under the influence of a vacuum bath. This is followed by a tomogram 
for a general spin-1 pure state
in Section \ref{sec:Tomogram-of-spin-1}. The tomograms of finite dimensional number-phase states 
under open quantum system evolution are discussed in Section \ref{sec:Tomogram-of-qutrit}. 
This is illustrated by a specific example of a three-level quantum (qutrit) system evolving
under a spontaneous emission channel. In  Section \ref{sec:Optical-tomogram}, 
we discuss the tomogram of an infinite dimensional system, the ubiquitous dissipative harmonic oscillator.
We conclude in Section \ref{sec:Conclusion}.

\section{Tomograms of single spin-$\frac{1}{2}$ states \label{sec:Tomograms-of-single-half}}

In this section, we study the tomograms for single spin-$\frac{1}{2}$
(qubit) atomic coherent state evolving under two general noise models, i.e., pure dephasing (QND)
and dissipative squeezed generalized amplitude damping (SGAD) evolution, incorporating the effects of dissipation, decoherence and bath squeezing.

\subsection{{\normalsize{}QND Evolution}}

The master equation of a quantum state under QND evolution \cite{QND}
is 
\begin{equation}
\begin{array}{lcl}
\dot{\rho}{}_{nm}^{s}\left(t\right) & = & \left[-\frac{i}{\hbar}\left(E_{n}-E_{m}\right)+i\dot{\eta}\left(t\right)\left(E_{n}^{2}-E_{m}^{2}\right)-\left(E_{n}-E_{m}\right)^{2}\dot{\gamma}\left(t\right)\right]\rho_{nm}^{s}\left(t\right),
\end{array}\label{eq:master-eq-QND}
\end{equation}
where $E_{n}$'s are the eigenvalues of the system Hamiltonian in
the system eigenbasis $|n\rangle$,
\[
\eta\left(t\right)=-\sum_{k}\frac{g_{k}^{2}}{\hbar^{2}\omega_{k}^{2}}\sin\left(\omega_{k}t\right),
\]
and 
\[
\begin{array}{lcl}
\gamma\left(t\right) & = & \frac{1}{2}\underset{k}{\sum}\frac{g_{k}^{2}}{\hbar^{2}\omega_{k}^{2}}\coth\left(\frac{\beta\hbar\omega_{k}}{2}\right)
\left|\left(e^{i\omega_{k}t}-1\right)\cosh\left(r_{k}\right)+\left(e^{-i\omega_{k}t}-1\right)\sinh\left(r_{k}\right)e^{2i\Phi_{k}}\right|^{2},
\end{array}
\]
 with $\beta=\frac{1}{k_{B}T}$. Here, $k_{B}$ is the Boltzmann constant,
$r_{k}$ and $\Phi_{k}$ are the bath squeezing parameters and $g_{k}$ is the system-bath coupling coefficient. The initial
density matrix for the atomic coherent state is given by
\begin{equation}
\rho^{s}\left(0\right)=|\alpha,\beta\rangle\langle\alpha,\beta|,\label{eq:initial-state-QND}
\end{equation}
where the atomic coherent state is given by 
\begin{equation}
\begin{array}{lcl}
|\alpha,\beta\rangle & = & \stackrel[m=-j]{j}{\sum}\left(\begin{array}{c}
2j\\
j+m
\end{array}\right)^{1/2}\sin\left(\frac{\alpha}{2}\right)^{j+m} \cos\left(\frac{\alpha}{2}\right)^{j-m}|j,m\rangle e^{-i\left(j+m\right)\beta}.
\end{array}\label{eq:atomic-coherent-state}
\end{equation}
The different elements of the density matrix in Eq. (\ref{eq:initial-state-QND})
at time $t$ under QND evolution becomes
\begin{equation}
\begin{array}{lcl}
\rho_{jm,jn}^{s}\left(t\right) & = & e^{-i\omega\left(m-n\right)t}e^{i\left(\hbar\omega\right)^{2}\left(m^{2}-n^{2}\right)\eta\left(t\right)} e^{-\left(\hbar\omega\right)^{2}\left(m-n\right)^{2}\gamma\left(t\right)}\rho_{jm,jn}^{s}\left(0\right),
\end{array}\label{eq:QND-densitymatirix}
\end{equation}
with $\rho_{jm,jn}^{s}\left(0\right)=\langle j,m|\rho^{s}\left(0\right)|j,n\rangle$.
Considering the initial state of the system as atomic coherent state, i.e., using Eq.
(\ref{eq:atomic-coherent-state}), different elements of the density
matrix in Eq. (\ref{eq:QND-densitymatirix}) at time $t=0$ are
\begin{equation}
\begin{array}{lcl}
\rho_{jm,jn}^{s}\left(0\right) & = & \left(\begin{array}{c}
2j\\
j+m
\end{array}\right)^{1/2}\left(\begin{array}{c}
2j\\
j+n
\end{array}\right)^{1/2}e^{i\left(n-m\right)\beta} \sin\left(\frac{\alpha}{2}\right)^{2j+m+n}\cos\left(\frac{\alpha}{2}\right)^{2j-m-n}.
\end{array}\label{eq:at_t0_with_acs}
\end{equation}
Using Eq. (\ref{eq:at_t0_with_acs}) as the initial density matrix
elements in Eq. (\ref{eq:QND-densitymatirix}), we can write all the
elements of the density matrix at time $t$ as
\begin{equation}
\begin{array}{lcl}
\rho_{jm,jn}^{s}\left(t\right) & = & \left(\begin{array}{c}
2j\\
j+m
\end{array}\right)^{1/2}\left(\begin{array}{c}
2j\\
j+n
\end{array}\right)^{1/2}e^{-i\omega\left(m-n\right)t} e^{i\left(\hbar\omega\right)^{2}\left(m^{2}-n^{2}\right)\eta\left(t\right)}\\
 & \times & e^{-\left(\hbar\omega\right)^{2}\left(m-n\right)^{2}\gamma\left(t\right)}\sin\left(\frac{\alpha}{2}\right)^{2j+m+n}\cos\left(\frac{\alpha}{2}\right)^{2j-m-n}e^{i\left(n-m\right)\beta}.
\end{array}\label{eq:at_t_with_acs}
\end{equation}

To obtain a tomogram of a spin-$\frac{1}{2}$ atomic coherent
state under QND evolution, we can express the density matrix in terms of
Wigner-Dicke states   as 
\begin{equation}
\rho^{\left(j\right)}\equiv\rho^{\left(j\right)}\left(t\right)=\sum_{m,m^{\prime}=-j}^{j}\rho_{m,m^{\prime}}^{\left(j\right)}|j,m\rangle\langle j,m^{\prime}|.\label{eq:wigner_dicke_state}
\end{equation}
The different elements of this density matrix $\rho_{m,m^{\prime}}^{\left(j\right)}=\langle m|\rho^{\left(j\right)}|m^{\prime}\rangle$
can be obtained using Eq. (\ref{eq:at_t_with_acs}), with $m,\,n\rightarrow m,\,m^{\prime},$
for $j=\frac{1}{2}$, $m,\,m^{\prime}=\pm\frac{1}{2}$. Subsequently,
the density matrix is obtained as

\begin{equation}
\rho^{\left(1/2\right)}=\left[\begin{array}{cc}
\sin^{2}\left(\frac{\alpha}{2}\right) & \frac{1}{2}e^{-i\omega t}e^{-\left(\hbar\omega\right)^{2}\gamma\left(t\right)}\sin\alpha e^{-i\beta}\\
\frac{1}{2}e^{i\omega t}e^{-\left(\hbar\omega\right)^{2}\gamma\left(t\right)}\sin\alpha e^{i\beta} & \cos^{2}\left(\frac{\alpha}{2}\right)
\end{array}\right].\label{eq:density-matrix-QND}
\end{equation}

\begin{figure}
\includegraphics[scale=0.6]{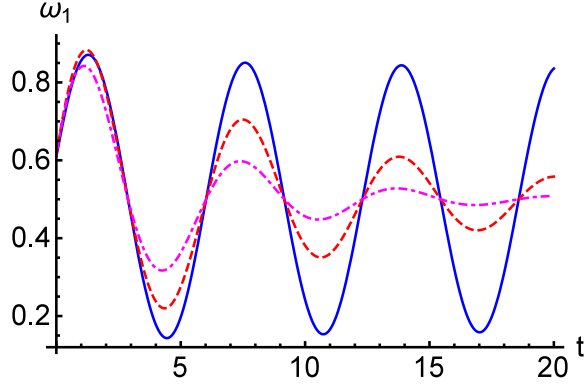}

\protect\caption{\label{fig:qnd-single}(Color online) The variation of the tomogram
with time ($t$) for single spin-$\frac{1}{2}$ atomic coherent state
in the presence of QND noise with bath parameters $\gamma_{0}=0.1,\,\omega_{c}=100,$
squeezing parameters $r=0,\,a=0,$ and $\omega=1.0$ and $\alpha=\frac{\pi}{2},\,\beta=\frac{\pi}{3},\,\widetilde{\beta}=\frac{\pi}{3},\,\widetilde{\gamma}=\frac{\pi}{4},$
in the units of $\hbar=k_{B}=1$. The smooth (blue) line, dashed (red)
line and dot-dashed (magenta) line correspond to the tomogram with
time for different temperatures $T=0,\,1$ and $2$, respectively.} 
\end{figure}

We can easily check that the trace of the density matrix $\left(\rho^{\left(1/2\right)}\right)$ is one, i.e.,
$\stackrel[m=-1/2]{1/2}{\sum}\rho_{m,m}^{\left(1/2\right)}=1.$ Further,
the tomogram of this state can be expressed as \cite{manko-spin-tom}
\begin{equation}
\begin{array}{lcl}
\omega\left(m_{1},\widetilde{\alpha},\widetilde{\beta},\widetilde{\gamma}\right) & = & \stackrel[m=-j]{j}{\sum}\stackrel[m^{\prime}=-j]{j}{\sum}D_{m_{1},m}^{\left(j\right)}\left(\widetilde{\alpha},\widetilde{\beta},\widetilde{\gamma}\right) \rho_{m,m^{\prime}}^{\left(j\right)}D_{m_{1},m^{\prime}}^{\left(j\right)*}\left(\widetilde{\alpha},\widetilde{\beta},\widetilde{\gamma}\right),
\end{array}\label{eq:tomogram}
\end{equation}
where $D_{m,m^{\prime}}^{\left(j\right)}\left(\widetilde{\alpha},\widetilde{\beta},\widetilde{\gamma}\right)$
is the Wigner $D$-function  
\begin{equation}
\begin{array}{lcl}
D_{m,m^{\prime}}^{\left(j\right)}\left(\widetilde{\alpha},\widetilde{\beta},\widetilde{\gamma}\right) & = & e^{-im\widetilde{\alpha}}d_{m,m^{\prime}}^{\left(j\right)}\left(\widetilde{\beta}\right)e^{-im^{\prime}\widetilde{\gamma}}\end{array}, \label{eq:wigner_D-function}
\end{equation}  and the notation used here is consistent with
that in Ref. \cite{varshalo}.
Here, $\widetilde{\alpha}$, $\widetilde{\beta}$, and $\widetilde{\gamma}$
are Euler angles $\equiv\phi$ , $\theta$, and $\psi$, with $\phi,\psi\in\left[0,2\pi\right]$,
and $\theta\in\left[0,\pi\right]$, and
\begin{equation}
\begin{array}{lcl}
d_{m,m^{\prime}}^{\left(j\right)}\left(\widetilde{\beta}\right) & = & \left[\frac{\left(j+m\right)!\left(j-m\right)!}{\left(j+m^{\prime}\right)!\left(j-m^{\prime}\right)!}\right]^{1/2}\left(\cos\frac{\widetilde{\beta}}{2}\right)^{m+m^{\prime}} \left(\sin\frac{\widetilde{\beta}}{2}\right)^{m-m^{\prime}}P_{j-m}^{\left(m-m^{\prime},m+m^{\prime}\right)}\left(\cos\widetilde{\beta}\right),
\end{array}\label{eq:D-function-d}
\end{equation}
where $P_{n}^{\left(a,b\right)}\left(x\right)$ are Jacobi polynomials.
A tomogram is the spin projection onto an arbitrary, rotated, axis.
The physical significance of the $D$-function is its connection to
the process of rotation and can be illustrated by
\begin{equation}
\begin{array}{lcl}
\langle j,m_{1}|R\left(\widetilde{\alpha},\widetilde{\beta},\widetilde{\gamma}\right)|j,m_{1}^{\prime}\rangle & = & D_{m_{1},m_{1}^{\prime}}^{\left(j\right)}\left(\widetilde{\alpha},\widetilde{\beta},\widetilde{\gamma}\right),\\
\langle j,m_{2}^{\prime}|R^{\dagger}\left(\widetilde{\alpha},\widetilde{\beta},\widetilde{\gamma}\right)|j,m_{1}\rangle & = & D_{m_{1},m_{2}^{\prime}}^{*\left(j\right)}\left(\widetilde{\alpha},\widetilde{\beta},\widetilde{\gamma}\right).
\end{array}\label{eq:Meaning-of-D}
\end{equation}
Here, $R\left(\widetilde{\alpha},\widetilde{\beta},\widetilde{\gamma}\right)$
stands for the operation of rotation about an axis whose orientation
is specified by $\widetilde{\alpha},\,\widetilde{\beta},$ and $\widetilde{\gamma}.$
Using the different values of $m$ and $m^{\prime}$, we can obtain
various Wigner $D$-functions as
\begin{equation}
\begin{array}{lcl}
D_{\frac{1}{2},-\frac{1}{2}}^{\left(1/2\right)}\left(\widetilde{\alpha},\widetilde{\beta},\widetilde{\gamma}\right) & = & -\sin\left(\frac{\widetilde{\beta}}{2}\right)e^{-\frac{i}{2}\left(\widetilde{\alpha}-\widetilde{\gamma}\right)},\\
D_{\frac{1}{2},\frac{1}{2}}^{\left(1/2\right)}\left(\widetilde{\alpha},\widetilde{\beta},\widetilde{\gamma}\right) & = & \cos\left(\frac{\widetilde{\beta}}{2}\right)e^{-\frac{i}{2}\left(\widetilde{\alpha}+\widetilde{\gamma}\right)},\\
D_{-\frac{1}{2},-\frac{1}{2}}^{\left(1/2\right)}\left(\widetilde{\alpha},\widetilde{\beta},\widetilde{\gamma}\right) & = & \cos\left(\frac{\widetilde{\beta}}{2}\right)e^{\frac{i}{2}\left(\widetilde{\alpha}+\widetilde{\gamma}\right)},\\
D_{-\frac{1}{2},\frac{1}{2}}^{\left(1/2\right)}\left(\widetilde{\alpha},\widetilde{\beta},\widetilde{\gamma}\right) & = & \sin\left(\frac{\widetilde{\beta}}{2}\right)e^{\frac{i}{2}\left(\widetilde{\alpha}-\widetilde{\gamma}\right)}.
\end{array}\label{eq:values-of-Ds}
\end{equation}
Using the first two relations of Eq. (\ref{eq:values-of-Ds})
and Eq. (\ref{eq:density-matrix-QND}), the first component of the
tomogram can be obtained from Eq. (\ref{eq:tomogram}) as 
\begin{equation}
\begin{array}{lcl}
\omega\left(\frac{1}{2},\widetilde{\alpha},\widetilde{\beta},\widetilde{\gamma}\right)\equiv\omega_{1} & = & \cos^{2}\left(\frac{\widetilde{\beta}}{2}\right)-\cos\widetilde{\beta}\cos^{2}\left(\frac{\alpha}{2}\right) - \frac{1}{2}\sin\widetilde{\beta}\sin\alpha\cos\left(\omega t+\beta+\widetilde{\gamma}\right) e^{-\left(\hbar\omega\right)^{2}\gamma\left(t\right)}.
\end{array}\label{eq:tomogram1-QND}
\end{equation}
From Eq. (\ref{eq:tomogram1-QND}), it can be inferred that the tomogram
is free from Euler angle $\widetilde{\alpha}$, and consequently 
is a function of  $\widetilde{\beta}$ and $\widetilde{\gamma}$ only,
or $f(\widetilde{\beta},\widetilde{\gamma})$. It is
worth mentioning here that $\widetilde{\gamma}$ and $\gamma\left(t\right)$
are two different parameters, the former being an Euler angle while the latter is responsible for decoherence. The variation of the tomogram is
given in Fig. \ref{fig:qnd-single} with time, for the different temperatures.
For the second component of the tomogram with $m_{1}=-\frac{1}{2}$,
using last two relations of Eq. (\ref{eq:values-of-Ds}) and substituting Eq.
(\ref{eq:density-matrix-QND}) in Eq. (\ref{eq:tomogram}), we obtain
\begin{equation}
\begin{array}{lcl}
\omega\left(-\frac{1}{2},\widetilde{\alpha},\widetilde{\beta},\widetilde{\gamma}\right)\equiv\omega_{2} & = & \cos^{2}\left(\frac{\widetilde{\beta}}{2}\right)-\cos\widetilde{\beta}\cos^{2}\left(\frac{\alpha}{2}\right) + \frac{1}{2}\sin\widetilde{\beta}\sin\alpha\cos\left(\omega t+\beta+\widetilde{\gamma}\right) e^{-\left(\hbar\omega\right)^{2}\gamma\left(t\right)}.
\end{array}\label{eq:tomogram-2-QND}
\end{equation}
We can check the validity of the tomogram obtained by
verifying that $\sum\omega_{i}=\begin{array}{lcl}
\omega_{1}+\omega_{2} & = & 1.\end{array}$ Interestingly, we can see  that the knowledge of one of the components
of the tomogram is enough to reconstruct the whole state.
Keeping this in mind, we have only shown the variation of $\omega_{1}$ in Fig. \ref{fig:qnd-single}
as $\omega_{2}=1-\omega_{1}$. 

\begin{figure}
\includegraphics[angle=-90,scale=0.65]{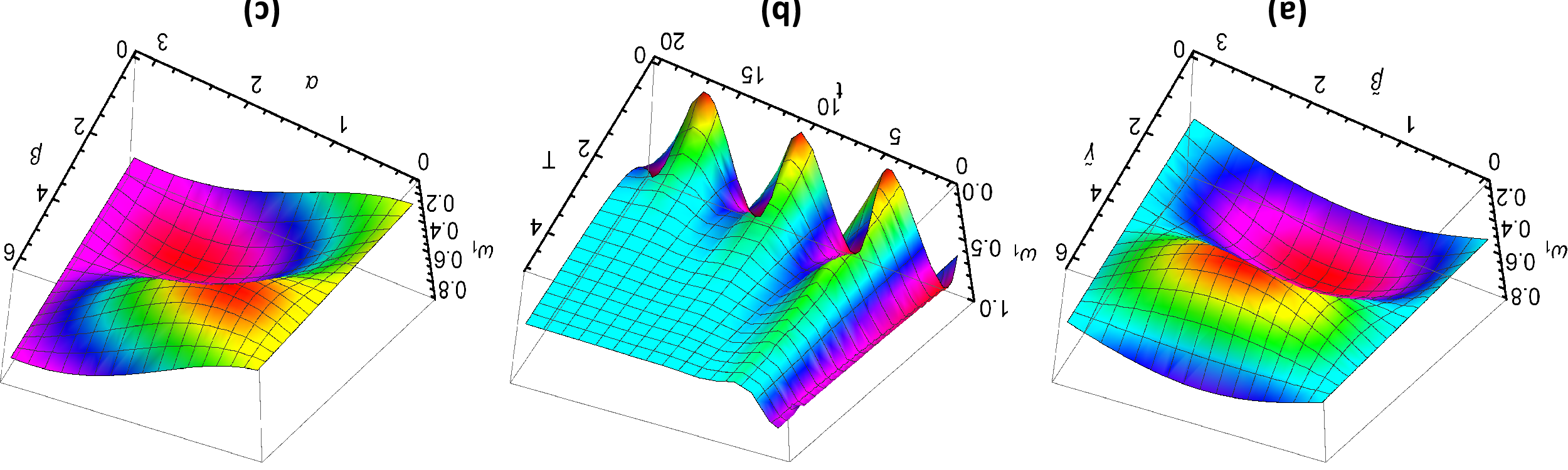}
\protect\caption{\label{fig:qnd-single-3D}(Color online) The dependence of the tomogram
on various parameters is depicted for single spin-$\frac{1}{2}$ atomic
coherent state in the presence of QND noise with bath parameters $\gamma_{0}=0.1,\,\omega_{c}=100,$
squeezing parameter $a=0,$ and $\omega=1.0$ in the units of $\hbar=k_{B}=1$.
In (a) the tomogram is shown as a function of $\widetilde{\beta}$
and $\widetilde{\gamma}$ with $\alpha=\frac{\pi}{2},\,\beta=\frac{\pi}{3},$
and $r=t=T=1$; while (b) exhibits the variation of the tomogram with
time and temperature for $r=0$ and $\alpha=\frac{\pi}{2},\,\beta=\frac{\pi}{3},\,\widetilde{\beta}=\frac{\pi}{3},\,\widetilde{\gamma}=\frac{\pi}{4}$.
Finally, the last plot (c) shows the dependence of the tomogram on
the atomic coherent state parameters $\alpha$ and $\beta$
with $\widetilde{\beta}=\frac{\pi}{3},\,\widetilde{\gamma}=\frac{\pi}{4}$ at time $t=1$
for bath squeezing parameter $r=1$ at $T=1$.}
\end{figure}

In Fig. \ref{fig:qnd-single}, we can easily see the expected behavior
of tomogram with increase in temperature for zero bath squeezing.  
Specifically, with increase in temperature, the tomogram tends to
randomize more quickly towards probability $1/2$. Fig. \ref{fig:qnd-single-3D}
further establishes the effect of the environment on the tomogram.
Particularly, Fig. \ref{fig:qnd-single-3D} b brings out the oscillatory nature
of tomogram with time while temperature tends to randomize it. Similarly,
Figs. \ref{fig:qnd-single-3D} a and c show the dependence of the tomogram
on Euler angles and the atomic coherent state parameters, respectively.

\subsection{{\normalsize{}Dissipative SGAD channel}}

Master equation for the dissipative evolution of a given state in
the squeezed generalized amplitude damping (SGAD) channel is given
by \cite{SGAD} 
\begin{equation}
\begin{array}{lcl}
\frac{d}{dt}\rho^{s}\left(t\right) & = & -\frac{i\omega}{2}\left[\sigma_{z},\rho^{s}\left(t\right)\right]+\gamma_{0}\left(N+1\right) \left\{ \sigma_{-}\rho^{s}\left(t\right)\sigma_{+}-\frac{1}{2}\sigma_{+}\sigma_{-}\rho^{s}\left(t\right)-\frac{1}{2}\rho^{s}\left(t\right)\sigma_{+}\sigma_{-}\right\} \\
 & + & \left\{ \sigma_{+}\rho^{s}\left(t\right)\sigma_{-}-\frac{1}{2}\sigma_{-}\sigma_{+}\rho^{s}\left(t\right)-\frac{1}{2}\rho^{s}\left(t\right)\sigma_{-}\sigma_{+}\right\} \gamma_{0}N-\gamma_{0}M\sigma_{+}\rho^{s}\left(t\right)\sigma_{+}-\gamma_{0}M^{*}\sigma_{-}\rho^{s}\left(0\right)\sigma_{-}.
\end{array}\label{eq:master_eq-SGAD}
\end{equation}
The density matrix for a quantum state under a dissipative SGAD channel
at time $t$ can be obtained, from the above equation, as 
\begin{equation}
\begin{array}{lcl}
\rho^{s}\left(t\right) & = & \frac{1}{4}\rho^{s}\left(0\right)f_{+}+\frac{1}{4}\sigma_{z}\rho^{s}\left(0\right)\sigma_{z}f_{-}-\frac{1}{4}\rho^{s}\left(0\right)\sigma_{z}g_{-} -  \frac{1}{4}\sigma_{z}\rho^{s}\left(0\right)g_{+}-\gamma_{0}\frac{\sinh\left(\alpha^{\prime}t\right)}{\alpha^{\prime}}e^{-\frac{\gamma^{\beta}t}{2}}\\
& \times & \left\{ M\sigma_{+}\rho^{s}\left(0\right)\sigma_{+}+M^{*}\sigma_{-}\rho^{s}\left(0\right)\sigma_{-}\right\} + \left(1-e^{-\gamma^{\beta}t}\right)\left\{ \frac{\gamma_{+}}{\gamma^{\beta}}\sigma_{-}\rho^{s}\left(0\right)\sigma_{+}+\frac{\gamma_{-}}{\gamma^{\beta}}\sigma_{+}\rho^{s}\left(0\right)\sigma_{-}\right\},
\end{array}\label{eq:density-matrix-SGAD}
\end{equation}
where $f_{\pm}=\left\{ 1+e^{-\gamma^{\beta}t} \pm 2\cosh\left(\alpha^{\prime}t\right)e^{-\frac{\gamma^{\beta}t}{2}} \right\},$ $g_{\pm}=\left\{ \frac{\gamma}{\gamma^{\beta}}\left(1-e^{-\gamma^{\beta}t} \right) \pm \frac{2i\omega}{\alpha^{\prime}}\sinh\left(\alpha^{\prime}t\right)e^{-\frac{\gamma^{\beta}t}{2}} \right\},$ $\gamma_{+}=\gamma_{0}\left(N+1\right)$, $\gamma_{-}=\gamma_{0}N$,
$\gamma^{\beta}=\gamma_{+}+\gamma_{-}$, $\gamma=\gamma_{+}-\gamma_{-}=\gamma_{0}$,
$\alpha^{\prime}=\sqrt{\gamma_{0}^{2}\left|M\right|^{2}-\omega^{2}}$;
and
\[
\begin{array}{lcl}
\sigma_{+} & = & |1\rangle\langle0|,\,\,\sigma_{-}=|0\rangle\langle1|,\\
\sigma_{z} & = & \sigma_{+}\sigma_{-}-\sigma_{-}\sigma_{+}\\
 & = & |1\rangle\langle1|-|0\rangle\langle0|\\
 & = & |e\rangle\langle e|-|g\rangle\langle g|.
\end{array}
\]
Also,
\[
\begin{array}{lclccc}
\sigma_{z}|g\rangle & = & -|g\rangle, & \sigma_{z}|e\rangle & = & |e\rangle;\\
\sigma_{+}|g\rangle & = & |e\rangle, & \sigma_{+}|e\rangle & = & 0;\\
\sigma_{-}|g\rangle & = & 0, & \sigma_{-}|e\rangle & = & |g\rangle.
\end{array}
\]
Here, $\gamma_{0}$ is the spontaneous emission rate, $M=-\frac{1}{2}\left\{ 2N_{th}+1\right\} \exp\left(i\phi\right)\sinh\left(2r\right)$,
and $N=N_{th}\left\{ \cosh^{2}\left(r\right)+\sinh^{2}\left(r\right)\right\} +\sinh^{2}\left(r\right),$
where $N_{th}=1/\left\{ \exp\left(\hbar\omega/k_{B}T\right)-1\right\} $
being the Planck distribution, and $r$ and the bath squeezing angle
($\phi$) are the bath squeezing parameters. The initial state, as
for the tomogram of a quantum state under QND evolution, is the atomic
coherent state given in Eq. (\ref{eq:initial-state-QND}). Using Eq.
(\ref{eq:density-matrix-SGAD}), the density matrix can be written
as
\begin{equation}
\rho^{s}\left(t\right)=\left[\begin{array}{cc}
\langle\frac{1}{2}|\rho^{s}\left(t\right)|\frac{1}{2}\rangle & \langle\frac{1}{2}|\rho^{s}\left(t\right)|-\frac{1}{2}\rangle\\
\langle-\frac{1}{2}|\rho^{s}\left(t\right)|\frac{1}{2}\rangle & \langle-\frac{1}{2}|\rho^{s}\left(t\right)|-\frac{1}{2}\rangle
\end{array}\right],\label{eq:final_density-matrix-SGAD}
\end{equation}
where the various terms are 
\[
\begin{array}{lcl}
\langle\frac{1}{2}|\rho^{s}\left(t\right)|\frac{1}{2}\rangle & = & \sin^{2}\left(\frac{\alpha}{2}\right)e^{-\gamma^{\beta}t}+\frac{\gamma_{-}}{\gamma^{\beta}}\left(1-e^{-\gamma^{\beta}t}\right),\\
\langle\frac{1}{2}|\rho^{s}\left(t\right)|-\frac{1}{2}\rangle & = & \frac{1}{2}\sin\alpha\left[\left\{ \cosh\left(\alpha^{\prime}t\right)-\frac{i\omega}{\alpha^{\prime}}\sinh\left(\alpha^{\prime}t\right)\right\} e^{-i\beta}-\frac{\gamma_{0}M}{\alpha^{\prime}}\sinh\left(\alpha^{\prime}t\right)e^{i\beta}\right]e^{-\frac{\gamma^{\beta}t}{2}},\\
\langle-\frac{1}{2}|\rho^{s}\left(t\right)|\frac{1}{2}\rangle & = & \frac{1}{2}\sin\alpha\left[\left\{ \cosh\left(\alpha^{\prime}t\right)+\frac{i\omega}{\alpha^{\prime}}\sinh\left(\alpha^{\prime}t\right)\right\} e^{i\beta}-\frac{\gamma_{0}M^{*}}{\alpha^{\prime}}\sinh\left(\alpha^{\prime}t\right)e^{-i\beta}\right]e^{-\frac{\gamma^{\beta}t}{2}},\\
\langle-\frac{1}{2}|\rho^{s}\left(t\right)|-\frac{1}{2}\rangle & = & \cos^{2}\left(\frac{\alpha}{2}\right)e^{-\gamma^{\beta}t}+\frac{\gamma_{+}}{\gamma^{\beta}}\left(1-e^{-\gamma^{\beta}t}\right),
\end{array}
\]
 and the density matrix can be seen to be normalized as
$\stackrel[m=-1/2]{1/2}{\sum}\langle m|\rho^{s}\left(t\right)|m\rangle=1.$

The tomogram of a state evolving in a dissipative SGAD channel, in analogy to the QND case, can be obtained 
by expanding the density matrix in the basis of the Wigner-Dicke states, as in Eq. (\ref{eq:wigner_dicke_state}). Using Eq. (\ref{eq:tomogram}), the first two relations 
of Eq. (\ref{eq:values-of-Ds}) and Eq. (\ref{eq:final_density-matrix-SGAD}), the first component of the tomogram is   

\begin{equation}
\begin{array}{lcl}
\omega\left(\frac{1}{2},\widetilde{\alpha},\widetilde{\beta},\widetilde{\gamma}\right)\equiv\omega_{1} & = & \sin^{2}\left(\frac{\widetilde{\beta}}{2}\right)\left\{ \cos^{2}\left(\frac{\alpha}{2}\right)e^{-\gamma^{\beta}t}+\frac{\gamma_{+}}{\gamma^{\beta}}\left(1-e^{-\gamma^{\beta}t}\right)\right\} +\cos^{2}\left(\frac{\widetilde{\beta}}{2}\right)\left\{ \sin^{2}\left(\frac{\alpha}{2}\right)e^{-\gamma^{\beta}t}+\frac{\gamma_{-}}{\gamma^{\beta}}\left(1-e^{-\gamma^{\beta}t}\right)\right\} \\
 & - & \frac{1}{2}\sin\widetilde{\beta}\left\{ e^{-i\widetilde{\gamma}}\left[\frac{1}{2}\sin\alpha e^{-i\beta}e^{-\frac{\gamma^{\beta}t}{2}}\left\{ \cosh\left(\alpha^{\prime}t\right)-\frac{i\omega}{\alpha^{\prime}}\sinh\left(\alpha^{\prime}t\right)\right\} -\frac{\gamma_{0}M}{2\alpha^{\prime}}\sin\alpha\sinh\left(\alpha^{\prime}t\right)e^{i\beta}e^{-\frac{\gamma^{\beta}t}{2}}\right]\right.\\
 & + & \left.{\rm c.c.}\right\} .
\end{array}\label{eq:tomogram1-SGAD}
\end{equation}

Again, we can check the validity of the analytic expression of the tomogram in the absence of
open system effects, i.e., by considering $\gamma_{0}=\gamma=0$, $\gamma_{+}=\gamma_{-}=0=\gamma^{\beta}$, which
leads to $\alpha^{\prime}=i\omega$, we have
\begin{equation}
\begin{array}{lcl}
\omega\left(\frac{1}{2},\widetilde{\alpha},\widetilde{\beta},\widetilde{\gamma}\right) & = & \cos^{2}\left(\frac{\widetilde{\beta}}{2}\right)-\cos\widetilde{\beta}\cos^{2}\left(\frac{\alpha}{2}\right) - \frac{1}{2}\sin\widetilde{\beta}\sin\alpha\cos\left(\omega t+\beta+\widetilde{\gamma}\right),
\end{array}\label{eq:check-tomogram1-SGAD}
\end{equation}
which is identical to the QND case, i.e., Eq. (\ref{eq:tomogram1-QND}),
with $\gamma\left(t\right)=0$. Similarly, using Eq. (\ref{eq:tomogram}),
the last two relations of Eq. (\ref{eq:values-of-Ds}), and Eq. (\ref{eq:final_density-matrix-SGAD}),
we obtain the second component as

\begin{equation}
\begin{array}{lcl}
\omega\left(-\frac{1}{2},\widetilde{\alpha},\widetilde{\beta},\widetilde{\gamma}\right)\equiv\omega_{2} & = & \cos^{2}\left(\frac{\widetilde{\beta}}{2}\right)\left\{ \cos^{2}\left(\frac{\alpha}{2}\right)e^{-\gamma^{\beta}t}+\frac{\gamma_{+}}{\gamma^{\beta}}\left(1-e^{-\gamma^{\beta}t}\right)\right\} +\sin^{2}\left(\frac{\widetilde{\beta}}{2}\right)\left\{ \sin^{2}\left(\frac{\alpha}{2}\right)e^{-\gamma^{\beta}t}+\frac{\gamma_{-}}{\gamma^{\beta}}\left(1-e^{-\gamma^{\beta}t}\right)\right\} \\
 & + & \frac{1}{2}\sin\widetilde{\beta}\left\{ e^{-i\widetilde{\gamma}}\left[\frac{1}{2}\sin\alpha e^{-i\beta}e^{-\frac{\gamma^{\beta}t}{2}}\left\{ \cosh\left(\alpha^{\prime}t\right)-\frac{i\omega}{\alpha^{\prime}}\sinh\left(\alpha^{\prime}t\right)\right\} -\frac{\gamma_{0}M}{2\alpha^{\prime}}\sin\alpha\sinh\left(\alpha^{\prime}t\right)e^{i\beta}e^{-\frac{\gamma^{\beta}t}{2}}\right]\right.\\
 & + & \left.{\rm c.c.}\right\} .
\end{array}\label{eq:tomogram2-SGAD}
\end{equation}

Similar to the first tomogram of the dissipative SGAD
channel, we can check the solution in the absence of the open system
effects which leads to $\alpha^{\prime}=i\omega$. This can be seen to be the same as the corresponding
QND case, i.e., Eq. (\ref{eq:tomogram-2-QND}), with $\gamma\left(t\right)=0$
\begin{equation}
\begin{array}{lcl}
\omega\left(-\frac{1}{2},\widetilde{\alpha},\widetilde{\beta},\widetilde{\gamma}\right) & = & \cos^{2}\left(\frac{\widetilde{\beta}}{2}\right)-\cos\widetilde{\beta}\sin^{2}\left(\frac{\alpha}{2}\right) + \frac{1}{2}\sin\widetilde{\beta}\sin\alpha\cos\left(\omega t+\beta+\widetilde{\gamma}\right).
\end{array}\label{eq:check-tomogram-2-SGAD}
\end{equation}
We can also check the validity of the tomogram as in the QND case
by $\sum\omega_{i}=\omega_{1}+\omega_{2}=1.$ Hence, as before, one component of the tomogram would be enough to recover
all the information. This is why, in the plots, we only show
the first component of the tomogram. The other component can be easily obtained from
it.

\begin{figure}
\includegraphics[scale=0.6]{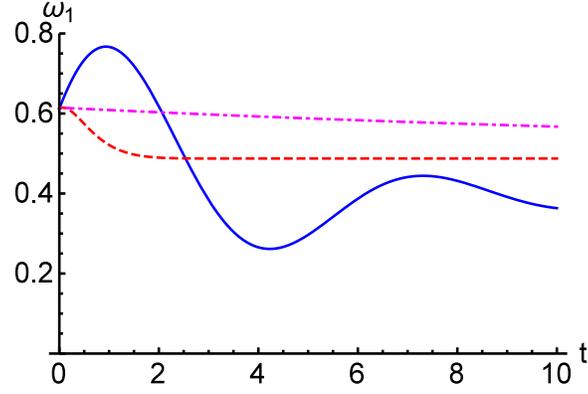}

\protect\caption{\label{fig:SGAD-ACS}(Color online) The tomogram varying
with time ($t$) is shown for a single spin-$\frac{1}{2}$
atomic coherent state in the presence of the SGAD noise for bath squeezing
angle $\phi=\pi$ in the units of $\hbar=k_{B}=1$, with
$\omega=1.0,\,\gamma_{0}=0.25,$ and $\alpha=\frac{\pi}{2},\,\beta=\frac{\pi}{3},\,\widetilde{\beta}=\frac{\pi}{3},\,\widetilde{\gamma}=\frac{\pi}{4}$.
The smooth (blue) line, dashed (red) line and dot-dashed (magenta)
line correspond to the tomogram with time for different temperatures
and squeezing parameters $T=1,\,10$ and $10$, and $r=0,\,0$ and
$1$, respectively.}
\end{figure}

\begin{figure}
\includegraphics[scale=0.55]{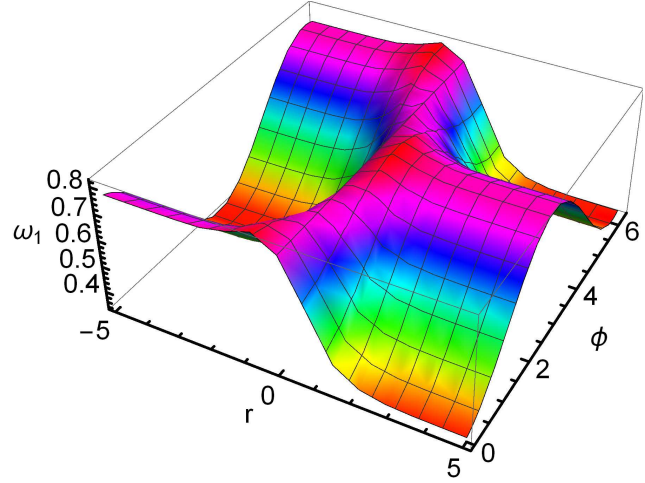}

\protect\caption{\label{fig:SGAD-3D-ACS}(Color online) The tomogram for single spin-$\frac{1}{2}$ atomic coherent
state in the presence of SGAD noise is
shown as a function of squeezing parameters $r$ and $\phi$ with
$\alpha=\frac{\pi}{2},\,\beta=\frac{\pi}{3},\,\widetilde{\beta}=\frac{\pi}{3},\,\widetilde{\gamma}=\frac{\pi}{4},$ and $\omega=1.0,\,\gamma_{0}=0.25,$
in the units of $\hbar=k_{B}=1$
for $T=1$ at time $t=1$.}
\end{figure}

The variation of tomogram with different parameters is shown in Figs.
\ref{fig:SGAD-ACS} and \ref{fig:SGAD-3D-ACS}. Fig.
\ref{fig:SGAD-ACS} exhibits the randomization of the tomogram with increase in
temperature.   This fact can be
observed in the smooth (blue) and dashed (red) lines. However, an interesting behavior is
observed here with respect to bath squeezing. It can be seen that
it takes relatively longer to randomize the tomogram
in presence of squeezing than in its absence, temperature remaining same, as illustrated by a comparison between the dot-dashed
(magenta) and dashed (red) line. 
This fact, in turn, establishes that
squeezing is a useful quantum resource. This behavior is
further elaborated in Fig. \ref{fig:SGAD-3D-ACS}, where the effect
of bath squeezing can be observed and is consistent with the
quadrature behavior of squeezing. This beneficial effect of squeezing is not observed for evolution under QND channel. From the present
analysis, it could be envisaged that a tomographic connection could be established between the state, under consideration and the generic open system interaction 
evolving it.    

\section{Tomogram of two spin-$\frac{1}{2}$ (qubit) states \label{sec:Tomogram-of-two-spin}}

Various two-qubit tomography schemes have been proposed in the recent past \cite{Manko-1/2,Manko-2spin,adam-2qubit-tom,comment12-ref1,comment12-ref2}. Specifically, the tomogram for two spin-$\frac{1}{2}$ (qubit) states can be obtained
using the star product scheme \cite{Manko-1/2,Manko-2spin}. In \cite{comment12-ref1}, two-qubit states were analyzed from the perspective of tomographic causal 
analysis, while in \cite{comment12-ref2}, an interesting connection between tomographic construction of two-qubit states to aspects of quantum correlations such as
discord and measurement induced disturbance was developed.   

For a two qubit state $\rho$ one can obtain the tomogram as 
\begin{equation}
\omega\left(m_{1},m_{2}\right)={\rm Tr}\left[\rho\left\{ Q_{1}\left(m_{1}\right)\otimes Q_{2}\left(m_{2}\right)\right\} \right],\label{eq:2qubitTomogram}
\end{equation}
where $Q_{i}\left(m_{i}\right)=U_{i}^{\dagger}\left|m_{i}\right\rangle \left\langle m_{i}\right|U_{i},$
and $m_{i}=\pm\frac{1}{2},$ while the unitary matrices $U_{i}$ are
\[
U_{i}=\left[\begin{array}{cc}
\cos\frac{\widetilde{\beta}_{i}}{2}\exp\left\{ \frac{i\left(\widetilde{\alpha}_{i}+\widetilde{\gamma}_{i}\right)}{2}\right\}  & \sin\frac{\widetilde{\beta}_{i}}{2}\exp\left\{ \frac{i\left(\widetilde{\alpha}_{i}-\widetilde{\gamma}_{i}\right)}{2}\right\} \\
-\sin\frac{\widetilde{\beta}_{i}}{2}\exp\left\{ -\frac{i\left(\widetilde{\alpha}_{i}-\widetilde{\gamma}_{i}\right)}{2}\right\}  & \cos\frac{\widetilde{\beta}_{i}}{2}\exp\left\{ -\frac{i\left(\widetilde{\alpha}_{i}+\widetilde{\gamma}_{i}\right)}{2}\right\} 
\end{array}\right]
\]
for $i\in\left\{ 1,2\right\}.$ Hence, the tomogram of the two qubit
state can be written as the diagonal elements of $\widetilde{\rho},$
where $\widetilde{\rho}=\left(U_{1}\otimes U_{2}\right)\rho\left(U_{1}\otimes U_{2}\right)^{\dagger}.$

\subsection{Tomogram of two qubits under dissipative evolution in a vacuum bath}

Now, we construct the tomogram of a two qubit state in a vacuum
bath under dissipative evolution, as discussed in Ref. \cite{squ-ther-bath}. The initial state of the system
is considered with one qubit in the excited state $\left|e_{1}\right\rangle $
and the other in the ground state $\left|g_{2}\right\rangle $, i.e.,
$\left|e_{1}\right\rangle \left|g_{2}\right\rangle $. The reduced
density matrix of the system of interest, here the two qubits, is 
\begin{equation}
\begin{array}{lcl}
\rho\left(t\right) & = & \left[\begin{array}{cccc}
\rho_{ee}\left(t\right) & \rho_{es}\left(t\right) & \rho_{ea}\left(t\right) & \rho_{eg}\left(t\right)\\
\rho_{es}^{*}\left(t\right) & \rho_{ss}\left(t\right) & \rho_{sa}\left(t\right) & \rho_{sg}\left(t\right)\\
\rho_{ea}^{*}\left(t\right) & \rho_{sa}^{*}\left(t\right) & \rho_{aa}\left(t\right) & \rho_{ag}\left(t\right)\\
\rho_{eg}^{*}\left(t\right) & \rho_{sg}^{*}\left(t\right) & \rho_{ag}^{*}\left(t\right) & \rho_{gg}\left(t\right)
\end{array}\right],\end{array}\label{eq:densitymatrix-vaccumbath}
\end{equation}
where the analytic form of all the elements of the density matrix is given in Appendix 1.

The tomogram can be thought of as a tomographic-probability vector
$\omega=\left[\omega_{1},\omega_{2},\omega_{3},\omega_{4}\right]^{T}$
(here $T$ corresponds to transpose of the vector), where each component
can be expressed analytically as 
\begin{equation}
\begin{array}{lcl}
\omega_{1}\left(t\right) & = & \frac{1}{4}\left[4\rho_{ee}\cos^{2}\frac{\widetilde{\beta}_{1}}{2}\cos^{2}\frac{\widetilde{\beta}_{2}}{2}+4\rho_{gg}\sin^{2}\frac{\widetilde{\beta}_{1}}{2}\sin^{2}\frac{\widetilde{\beta}_{2}}{2}\right. + \left(\rho_{aa}+\rho_{ss}\right)\left(1-\cos\widetilde{\beta}_{1}\cos\widetilde{\beta}_{2}\right)\\
 & - & \left(\rho_{aa}-\rho_{ss}\right)\sin\widetilde{\beta}_{1}\sin\widetilde{\beta}_{2}\cos\left(\widetilde{\gamma}_{1}-\widetilde{\gamma}_{2}\right)
+ \left\{ \rho_{sa}\left(\cos\widetilde{\beta}_{1}-\cos\widetilde{\beta}_{2}-i\sin\widetilde{\beta}_{1}\sin\widetilde{\beta}_{2}\right.\sin\left(\widetilde{\gamma}_{1}-\widetilde{\gamma}_{2}\right)\right)\\
 & + & \sqrt{2}\left[\left(\left(-\rho_{ea}+\rho_{es}\right)\cos^{2}\frac{\widetilde{\beta}_{2}}{2}+\left(\rho_{ag}+\rho_{sg}\right)\sin^{2}\frac{\widetilde{\beta}_{2}}{2}\right)\right.\sin\widetilde{\beta}_{1}\exp\left(i\widetilde{\gamma}_{1}\right)+\sin\widetilde{\beta}_{2}\exp\left(i\widetilde{\gamma}_{2}\right)\\
 & \times & \left.\left(\left(\rho_{ea}+\rho_{es}\right)\cos^{2}\frac{\widetilde{\beta}_{1}}{2}-\left(\rho_{ag}-\rho_{sg}\right)\sin^{2}\frac{\widetilde{\beta}_{1}}{2}\right)\right]+ \left.\left.\exp\left(i\widetilde{\gamma}_{1}+i\widetilde{\gamma}_{2}\right)\rho_{eg}\sin\widetilde{\beta}_{1}\sin\widetilde{\beta}_{2}+{\rm c.c.}\right\} \right],
\end{array}\label{eq:2qubit-tomo1}
\end{equation}
\begin{equation}
\begin{array}{lcl}
\omega_{2}\left(t\right) & = & \frac{1}{4}\left[4\rho_{ee}\cos^{2}\frac{\widetilde{\beta}_{1}}{2}\sin^{2}\frac{\widetilde{\beta}_{2}}{2}+4\rho_{gg}\sin^{2}\frac{\widetilde{\beta}_{1}}{2}\cos^{2}\frac{\widetilde{\beta}_{2}}{2}\right. + \left(\rho_{aa}+\rho_{ss}\right)\left(1+\cos\widetilde{\beta}_{1}\cos\widetilde{\beta}_{2}\right)\\
 & + & \left(\rho_{aa}-\rho_{ss}\right)\sin\widetilde{\beta}_{1}\sin\widetilde{\beta}_{2}\cos\left(\widetilde{\gamma}_{1}-\widetilde{\gamma}_{2}\right) +\left\{ \rho_{sa}\left(\cos\widetilde{\beta}_{1}+\cos\widetilde{\beta}_{2}+i\sin\widetilde{\beta}_{1}\sin\widetilde{\beta}_{2}\right.\sin\left(\widetilde{\gamma}_{1}-\widetilde{\gamma}_{2}\right)\right)\\
 & + & \sqrt{2}\left[\left(\left(\rho_{ag}+\rho_{sg}\right)\cos^{2}\frac{\widetilde{\beta}_{2}}{2}-\left(\rho_{ea}-\rho_{es}\right)\sin^{2}\frac{\widetilde{\beta}_{2}}{2}\right)\sin\widetilde{\beta}_{1}\exp\left(i\widetilde{\gamma}_{1}\right)\right.\\
 & + & \sin\widetilde{\beta}_{2}\exp\left(i\widetilde{\gamma}_{2}\right)\left.\left(-\left(\rho_{ea}+\rho_{es}\right)\cos^{2}\frac{\widetilde{\beta}_{1}}{2}+\left(\rho_{ag}-\rho_{sg}\right)\sin^{2}\frac{\widetilde{\beta}_{1}}{2}\right)\right] - \left.\left.\exp\left(i\widetilde{\gamma}_{1}+i\widetilde{\gamma}_{2}\right)\rho_{eg}\sin\widetilde{\beta}_{1}\sin\widetilde{\beta}_{2}+{\rm c.c.}\right\} \right],
\end{array}\label{eq:2qubit-tomo2}
\end{equation}
\begin{equation}
\begin{array}{lcl}
\omega_{3}\left(t\right) & = & \frac{1}{4}\left[4\rho_{ee}\sin^{2}\frac{\widetilde{\beta}_{1}}{2}\cos^{2}\frac{\widetilde{\beta}_{2}}{2}+4\rho_{gg}\cos^{2}\frac{\widetilde{\beta}_{1}}{2}\sin^{2}\frac{\widetilde{\beta}_{2}}{2}\right.
 + \left(\rho_{aa}+\rho_{ss}\right)\left(1+\cos\widetilde{\beta}_{1}\cos\widetilde{\beta}_{2}\right)\\
 & + & \left(\rho_{aa}-\rho_{ss}\right)\sin\widetilde{\beta}_{1}\sin\widetilde{\beta}_{2}\cos\left(\widetilde{\gamma}_{1}-\widetilde{\gamma}_{2}\right)
+ \left\{ -\rho_{sa}\left(\cos\widetilde{\beta}_{1}+\cos\widetilde{\beta}_{2}-i\sin\widetilde{\beta}_{1}\sin\widetilde{\beta}_{2}\right.\sin\left(\widetilde{\gamma}_{1}-\widetilde{\gamma}_{2}\right)\right)\\
 & + & \sqrt{2}\left[\left(-\left(\rho_{ag}+\rho_{sg}\right)\sin^{2}\frac{\widetilde{\beta}_{2}}{2}+\left(\rho_{ea}-\rho_{es}\right)\cos^{2}\frac{\widetilde{\beta}_{2}}{2}\right)\sin\widetilde{\beta}_{1}\exp\left(i\widetilde{\gamma}_{1}\right)\right.\\
 & + & \sin\widetilde{\beta}_{2}\exp\left(i\widetilde{\gamma}_{2}\right) \left.\left(\left(\rho_{ea}+\rho_{es}\right)\sin^{2}\frac{\widetilde{\beta}_{1}}{2}-\left(\rho_{ag}-\rho_{sg}\right)\cos^{2}\frac{\widetilde{\beta}_{1}}{2}\right)\right]
 - \left.\left.\exp\left(i\widetilde{\gamma}_{1}+i\widetilde{\gamma}_{2}\right)\rho_{eg}\sin\widetilde{\beta}_{1}\sin\widetilde{\beta}_{2}+{\rm c.c.}\right\} \right],
\end{array}\label{eq:2qubit-tomo3}
\end{equation}

\begin{figure}
\begin{centering}
\includegraphics[angle=-90,scale=0.9]{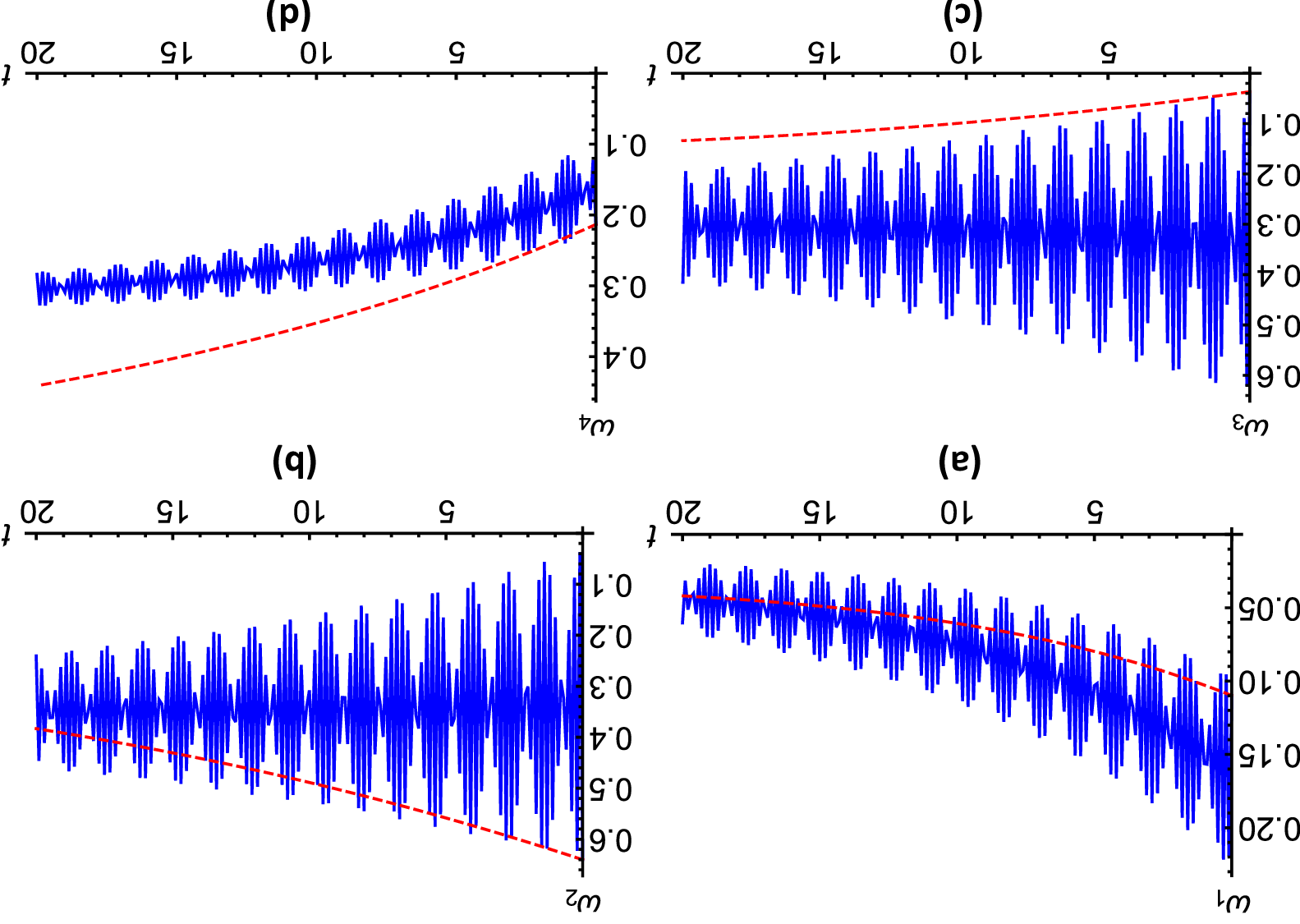}

\protect\caption{\label{fig:Vacuum-bath}(Color online) Various components
of the tomogram changing with time are shown in (a)-(d) for the two-qubit state,
in the presence of vacuum bath, with $\widetilde{\beta}_{1}=\frac{\pi}{3},\,\widetilde{\beta}_{2}=\frac{\pi}{4},\,\widetilde{\gamma}_{1}=\frac{\pi}{3},\,\widetilde{\gamma}_{2}=\frac{\pi}{4}$
and the inter-qubit spacing $r_{12}=0.05$ (2.0) corresponding to smooth blue (red dashed) line.}
\end{centering}
\end{figure}

and 
\begin{equation}
\begin{array}{lcl}
\omega_{4}\left(t\right) & = & \frac{1}{4}\left[4\rho_{ee}\sin^{2}\frac{\widetilde{\beta}_{1}}{2}\sin^{2}\frac{\widetilde{\beta}_{2}}{2}+4\rho_{gg}\cos^{2}\frac{\widetilde{\beta}_{1}}{2}\cos^{2}\frac{\widetilde{\beta}_{2}}{2}\right. +  \left(\rho_{aa}+\rho_{ss}\right)\left(1-\cos\widetilde{\beta}_{1}\cos\widetilde{\beta}_{2}\right)\\
 & - & \left(\rho_{aa}-\rho_{ss}\right)\sin\widetilde{\beta}_{1}\sin\widetilde{\beta}_{2}\cos\left(\widetilde{\gamma}_{1}-\widetilde{\gamma}_{2}\right)
+ \left\{ -\rho_{sa}\left(\cos\widetilde{\beta}_{1}-\cos\widetilde{\beta}_{2}+i\sin\widetilde{\beta}_{1}\sin\widetilde{\beta}_{2}\right.\sin\left(\widetilde{\gamma}_{1}-\widetilde{\gamma}_{2}\right)\right)\\
 & + & \sqrt{2}\left[\left(-\left(\rho_{ag}+\rho_{sg}\right)\cos^{2}\frac{\widetilde{\beta}_{2}}{2}+\left(\rho_{ea}-\rho_{es}\right)\sin^{2}\frac{\widetilde{\beta}_{2}}{2}\right)\sin\widetilde{\beta}_{1}\exp\left(i\widetilde{\gamma}_{1}\right)\right.\\
 & + & \sin\widetilde{\beta}_{2}\exp\left(i\widetilde{\gamma}_{2}\right)\left.\left(-\left(\rho_{ea}+\rho_{es}\right)\sin^{2}\frac{\widetilde{\beta}_{1}}{2}+\left(\rho_{ag}-\rho_{sg}\right)\cos^{2}\frac{\widetilde{\beta}_{1}}{2}\right)\right] - \left.\left.\exp\left(i\widetilde{\gamma}_{1}+i\widetilde{\gamma}_{2}\right)\rho_{eg}\sin\widetilde{\beta}_{1}\sin\widetilde{\beta}_{2}+{\rm c.c.}\right\} \right].
\end{array}\label{eq:2qubit-tomo4}
\end{equation}
Here, $\rho_{ij}$ are the elements of the matrix in Eq. (\ref{eq:densitymatrix-vaccumbath})
and are given in Appendix 1.
For simplicity of notations the time dependence in the arguments
of matrix elements is omitted. Similar to the tomograms for single
spin-$\frac{1}{2}$ states the tomogram obtained here is also free
from $\widetilde{\alpha}.$

As in the cases of single qubit tomograms, we can again verify that
the tomogram obtained here satisfies the condition $\sum\omega_{i}=\rho_{ee}+\rho_{gg}+\rho_{aa}+\rho_{ss},$
which is the trace of the density matrix given in Eq. (\ref{eq:densitymatrix-vaccumbath}),
and hence equal to one.

For the case of identical qubits considered here, we take the wave-vector
and mean frequency to be $k_{0}=\omega_{0}=1$, the spontaneous emission rate
$\Gamma_{j}=0.05$ and $\hat{\mu}\cdot\hat{r}_{ij}=0$.  
Here, $\hat{\mu}$ is the unit vector along the atomic transition
dipole moment and $\hat{r}_{ij}$ is the inter-atomic distance. Further,
the initial state of the system is taken to be   
$\rho_{ee}\left(0\right)=\rho_{gg}\left(0\right)=\rho_{es}\left(0\right)=\rho_{ea}\left(0\right)=\rho_{eg}\left(0\right)=\rho_{sg}\left(0\right)=
\rho_{ag}\left(0\right)=0$,
and $\rho_{ss}\left(0\right)=\rho_{aa}\left(0\right)=\rho_{sa}\left(0\right)=0.5$.

\begin{figure}
\begin{centering}
\includegraphics[angle=-90,scale=0.9]{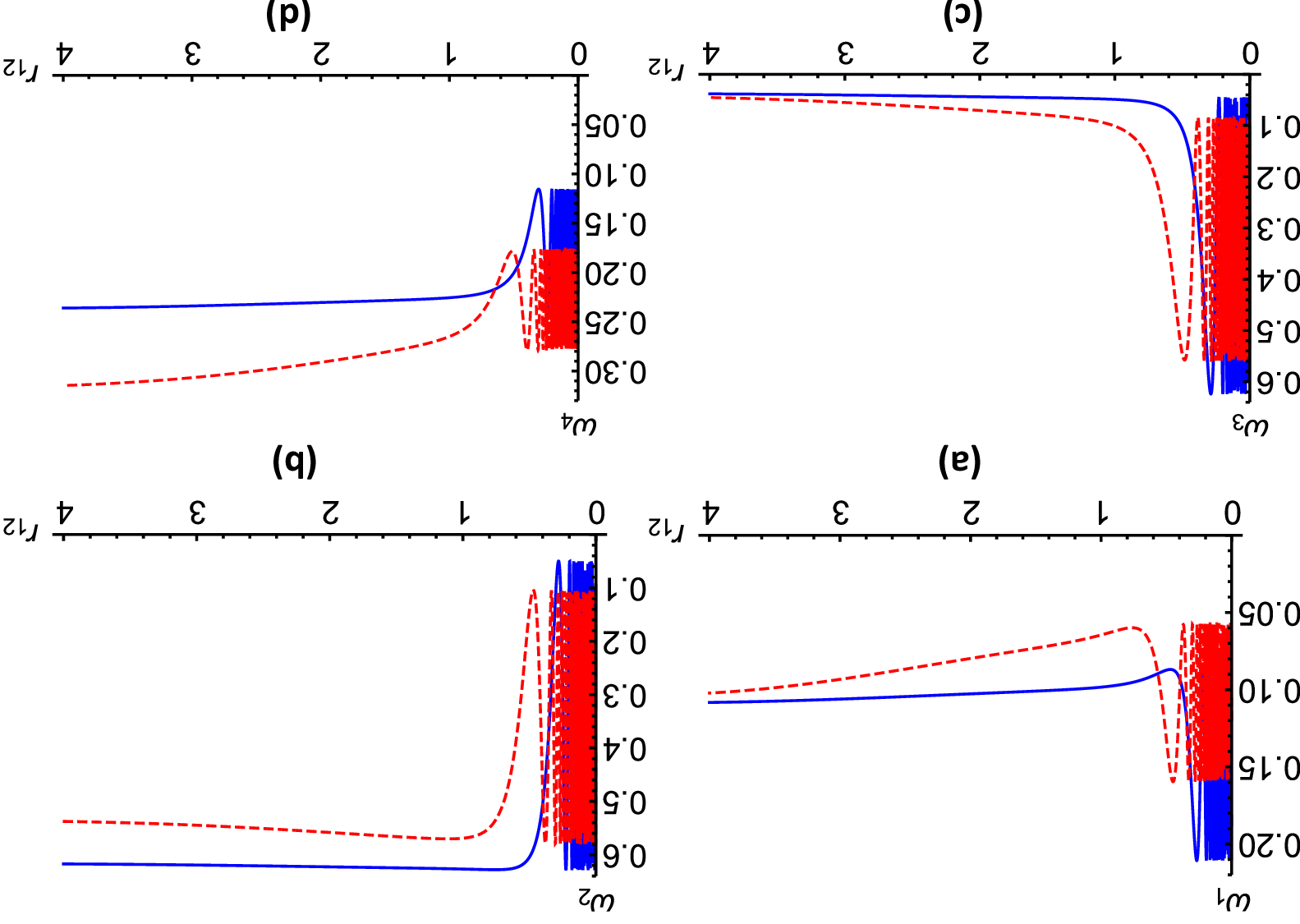}

\protect\caption{\label{fig:Vacuum-bath-r}(Color online) (a)-(d) depict the tomogram
for the two qubit state, interacting with a vacuum bath, as a function
of the inter-qubit spacing at $t=1$ (smooth blue line), and $t=5$
(red dashed line). For all the plots $\widetilde{\beta}_{1}=\frac{\pi}{3},\,\widetilde{\beta}_{2}=\frac{\pi}{4},\,\widetilde{\gamma}_{1}=\frac{\pi}{3},\,\widetilde{\gamma}_{2}=\frac{\pi}{4}$. }
\end{centering}
\end{figure}

The variation of all four components of the tomogram is shown with
different parameters in Figs. \ref{fig:Vacuum-bath} and \ref{fig:Vacuum-bath-r}.
In Fig. \ref{fig:Vacuum-bath}, large   oscillations can be observed
for small interqubit spacing, which is consistent with the earlier
observations in a plethora of scenario \cite{our-qd-paper,squ-ther-bath,GP}.
Fig. \ref{fig:Vacuum-bath-r} further demonstrates similar behavior for small interqubit spacing. For small
interqubit spacing the ambient environment opens up a channel between the qubits resulting in enhancement of oscillations.

\section{Tomogram of single spin-1 state \label{sec:Tomogram-of-spin-1}}

The tomograms for finite spin states have been considered, among others, in Refs. \cite{manko-spin-tom,spin-j-tom,manko-spin, comment6-ref1}. 
In continuation with the theme of this work, we take up an arbitrary spin-1 
state  
\begin{equation}
\begin{array}{lcl}
\psi^{\left(1\right)} & = & N\left[\begin{array}{c}
a\\
b\\
c
\end{array}\right]\end{array},\label{eq:spin-1_state}
\end{equation}

where $N=\frac{1}{\sqrt{\left|a\right|^{2}+\left|b\right|^{2}+\left|c\right|^{2}}}$
is the normalization factor. The corresponding density matrix is 
\begin{equation}
\begin{array}{lcl}
\rho^{\left(1\right)} & = & \left|N\right|^{2}\left[\begin{array}{ccc}
\left|a\right|^{2} & ab^{*} & ac^{*}\\
a^{*}b & \left|b\right|^{2} & bc^{*}\\
a^{*}c & b^{*}c & \left|c\right|^{2}
\end{array}\right]\end{array}.\label{eq:density_mat-spin-1}
\end{equation}
Here, we restrict ourselves to obtaining the tomogram for the state (\ref{eq:density_mat-spin-1}), without considering open system effects. Using Eqs. (\ref{eq:tomogram})-(\ref{eq:D-function-d}), all the Wigner
$D$-functions for the tomogram can be calculated as before.
Using Eqs. (\ref{eq:tomogram}) and (\ref{eq:density_mat-spin-1}),
$\omega\left(1,\widetilde{\alpha},\widetilde{\beta},\widetilde{\gamma}\right)$
can be written as

\begin{equation}
\begin{array}{lcl}
\omega\left(1,\widetilde{\alpha},\widetilde{\beta},\widetilde{\gamma}\right)\equiv\omega_{1} & = & \left|N\right|^{2}\left[\left\{ \frac{\left|a\right|^{2}}{4}\left(1+\cos\widetilde{\beta}\right)^{2}+\frac{\left|b\right|^{2}}{2}\sin^{2}\widetilde{\beta}+\frac{\left|c\right|^{2}}{4}\left(1-\cos\widetilde{\beta}\right)^{2}\right\} \right.\\
 & + & \left.\left\{ \left(-\frac{ab^{*}e^{i\widetilde{\gamma}}}{2\sqrt{2}}\right)\sin\widetilde{\beta}\left(1+\cos\widetilde{\beta}\right)+\frac{ac^{*}e^{2i\widetilde{\gamma}}}{4}\sin^{2}\widetilde{\beta}-\frac{bc^{*}e^{i\widetilde{\gamma}}}{2\sqrt{2}}\sin\widetilde{\beta}\left(1-\cos\widetilde{\beta}\right)+{\rm c.c.}\right\} \right].
\end{array}\label{eq:spin-1_tomgram1}
\end{equation}
Similarly, $\omega\left(0,\widetilde{\alpha},\widetilde{\beta},\widetilde{\gamma}\right)$
can be obtained as

\begin{equation}
\begin{array}{lcl}
\omega\left(0,\widetilde{\alpha},\widetilde{\beta},\widetilde{\gamma}\right) & \equiv\omega_{0}= & \left|N\right|^{2}\left[\left\{ \frac{\left|a\right|^{2}}{2}\sin^{2}\widetilde{\beta}+\left|b\right|^{2}\cos^{2}\widetilde{\beta}+\frac{\left|c\right|^{2}}{2}\sin^{2}\widetilde{\beta}\right\} \right.\\
 & + & \left.\left\{ \frac{ab^{*}e^{i\widetilde{\gamma}}}{2\sqrt{2}}\sin2\widetilde{\beta}-\frac{ac^{*}e^{2i\widetilde{\gamma}}}{2}\sin^{2}\widetilde{\beta}-\frac{bc^{*}e^{i\widetilde{\gamma}}}{2\sqrt{2}}\sin2\widetilde{\beta}+{\rm c.c.}\right\} \right].
\end{array}\label{eq:spin-1_tomogram2}
\end{equation}
Also, $\omega\left(-1,\widetilde{\alpha},\widetilde{\beta},\widetilde{\gamma}\right)$
is

\begin{equation}
\begin{array}{lcl}
\omega\left(-1,\widetilde{\alpha},\widetilde{\beta},\widetilde{\gamma}\right)\equiv\omega_{-1} & = & \left|N\right|^{2}\left[\left\{ \frac{\left|a\right|^{2}}{4}\left(1-\cos\widetilde{\beta}\right)^{2}+\frac{\left|b\right|^{2}}{2}\sin^{2}\widetilde{\beta}+\frac{\left|c\right|^{2}}{4}\left(1+\cos\widetilde{\beta}\right)^{2}\right\} \right.\\
 & + & \left.\left\{ \frac{ab^{*}e^{i\widetilde{\gamma}}}{2\sqrt{2}}\sin\widetilde{\beta}\left(1-\cos\widetilde{\beta}\right)+\frac{ac^{*}e^{2i\widetilde{\gamma}}}{4}\sin^{2}\widetilde{\beta}+\frac{bc^{*}e^{i\widetilde{\gamma}}}{2\sqrt{2}}\sin\widetilde{\beta}\left(1+\cos\widetilde{\beta}\right)+{\rm c.c.}\right\} \right].
\end{array}\label{eq:spin-1_tomogram3}
\end{equation}

Interestingly, the tomogram obtained for a general
spin-1 quantum state is also free from $\widetilde{\alpha}$ as
for spin-$\frac{1}{2}$ cases discussed above.
Further, it can be checked here that the tomogram satisfies the condition
$\omega_{1}+\omega_{0}+\begin{array}{lcl}
\omega_{-1} & = & 1.\end{array}$ For $a=1,b=0=c$, the tomogram, obtained here, is
seen to be consistent with the results reported earlier \cite{manko-spin},
\begin{equation}
\begin{array}{lcl}
\omega\left(1,\widetilde{\alpha},\widetilde{\beta},\widetilde{\gamma}\right) & = & \frac{\left(1+\cos\widetilde{\beta}\right)^{2}}{4},\\
\omega\left(0,\widetilde{\alpha},\widetilde{\beta},\widetilde{\gamma}\right) & = & \frac{\left(1-\cos^{2}\widetilde{\beta}\right)}{2},\\
\omega\left(-1,\widetilde{\alpha},\widetilde{\beta},\widetilde{\gamma}\right) & = & \frac{\left(1-\cos\widetilde{\beta}\right)^{2}}{4}.
\end{array}\label{eq:Manko's-spin-1}
\end{equation}

\begin{figure}
\includegraphics[angle=-90,scale=0.65]{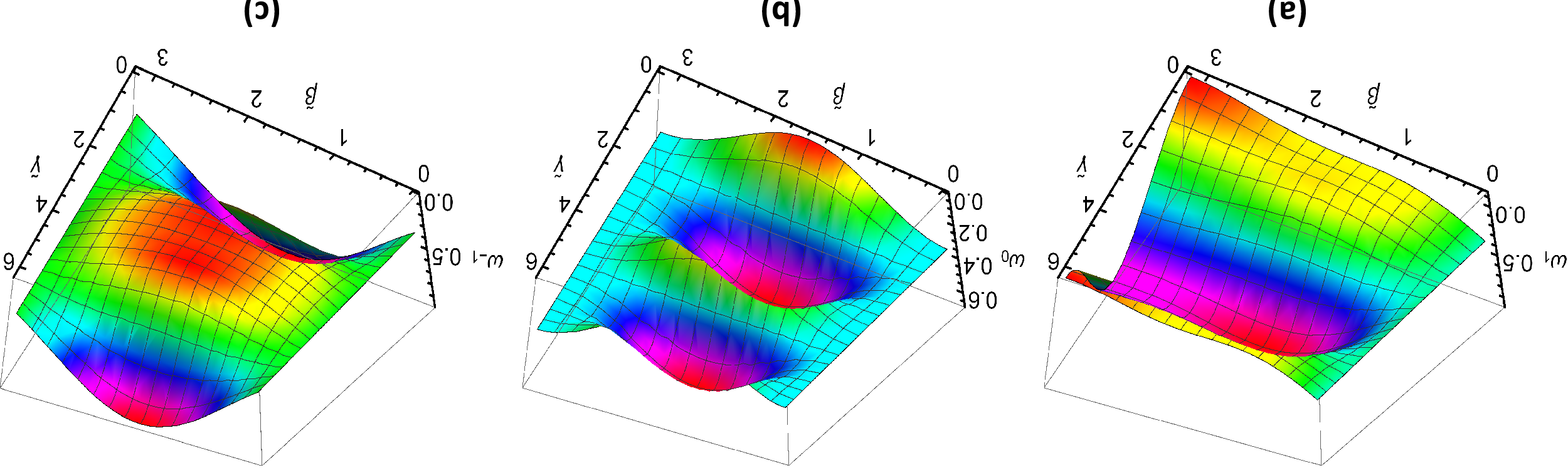}

\protect\caption{\label{fig:spin1-3D}(Color online) The dependence of all three components
of tomogram for single spin-1 state with Euler angles for $a=b=c=\frac{1}{\sqrt{3}}$. }
\end{figure}

The variation of all three components of the tomogram with Euler angles
is given in Fig. \ref{fig:spin1-3D}. The peaks in one component have
corresponding valleys in other components of the tomogram, which are
manifestations of normalization of the tomogram to one.

\section{Tomogram of a finite dimensional state \label{sec:Tomogram-of-qutrit}}

In this section, we discuss tomography of finite dimensional states.
The tomogram of a finite dimensional state can be defined as \cite{fss-tom}
\begin{equation}
\omega\left(m,t,q\right)=\stackrel[\chi=0]{d-1}{\sum}W\left(tm-q\chi,qm+t\chi\right),\label{eq:fss-tom}
\end{equation}
where $W\left(\chi,m\right)$ is the discrete Wigner function with
integers $t$ and $q$, depicting complementarity between number $m$ (denoting angular momentum) and phase $\chi$ of finite dimensional states. Here, the discrete Wigner function is expressed
as 
\begin{equation}
W\left(\chi,m\right)=\frac{1}{d}\stackrel[\Theta=0]{d-1}{\sum}\exp\left(\frac{4\pi i}{d}m\Theta\right)\left\langle \chi-\Theta\right|\rho\left|\chi+\Theta\right\rangle .\label{eq:wig-fss}
\end{equation}
It would be apt here to mention that the phase states $\left|\chi\right\rangle $
are periodic, such that $\left|\chi+d\right\rangle =\left|\chi\right\rangle $.
A general $d$ dimensional density matrix can be written in Weyl operator
basis as \cite{gen-bloch-vector} 
\begin{equation}
\rho=\frac{1}{d}\mathbb{I}+\stackrel[n,m=0]{d-1}{\sum}b_{nm}U_{nm}\label{eq:rho-weyl-op}
\end{equation}
 with $b_{00}=0$ and $U_{nm}=\stackrel[\alpha=0]{d-1}{\sum}\exp\left(\frac{2\pi i}{d}{\alpha}n\right)\left|\alpha\right\rangle \left\langle \alpha+m\right|$.
We make use of the periodicity of phase states in computations involving $U_{nm}$.

Up to now the treatment is applicable to any generic finite dimensional system. 
Here, for concreteness, we concentrate on an important finite dimensional system, viz. a qutrit with $d=3$. We study the effect of spontaneous emission (SE) channel \cite{noisy-qutrit} on the qutrit. 
SE is a dissipative process which can be modeled by the following Kraus operators
\[
\begin{array}{lcl}
K_{0} & = & \left[\begin{array}{ccc}
1 & 0 & 0\\
0 & e^{-\frac{\eta_{1}t}{2}} & 0\\
0 & 0 & e^{-\frac{\eta_{2}t}{2}}
\end{array}\right],\end{array}
\]
\[
\begin{array}{lcl}
K_{1} & = & \left[\begin{array}{ccc}
0 & \sqrt{1-e^{-\eta_{1}t}} & 0\\
0 & 0 & 0\\
0 & 0 & 0
\end{array}\right],\end{array}
\]
\begin{equation}
\begin{array}{lcl}
K_{2} & = & \left[\begin{array}{ccc}
0 & 0 & \sqrt{1-e^{-\eta_{2}t}}\\
0 & 0 & 0\\
0 & 0 & 0
\end{array}\right],\end{array}\label{eq:krauss-spon}
\end{equation}
where $\eta_{1}$ and $\eta_{2}$ are two Einstein coefficients which control the
population of the excited states. Thus, we can write the density matrix
of an arbitrary three dimensional state at time $t$ evolving under the spontaneous
emission channel as
\begin{equation}
\rho\left(t\right)=\stackrel[j,k=0]{2}{\sum}A_{jk}\left(t\right)\left|j\right\rangle \left\langle k\right|=\stackrel[i=0]{2}{\sum}K_{i}\rho\left(0\right)K_{i}^{\dagger}.\label{eq:rhot-qutrit}
\end{equation}
Using this we can obtain 
\begin{equation}
W\left(\chi,m,t\right)=\frac{1}{3}\stackrel[\Theta=0]{2}{\sum}A_{\chi-\Theta,\chi+\Theta}\left(t\right)\exp\left(\frac{4\pi i}{3}m\Theta\right).\label{eq:wig-qutrit}
\end{equation}
As mentioned above $\chi-\Theta$ and $\chi+\Theta$ are mod $d$ operations. Here, we have considered an initial density matrix given by Eq. (\ref{eq:rho-weyl-op})
with $b_{01}=b_{10}=\frac{1}{4}$ and $b_{12}=b_{21}=\frac{1}{5}$,
while the remaining coefficients can be obtained from these values.
Hence, the obtained tomogram with $t=0$ and $q=1$ has three
components as 
\begin{equation}
\begin{array}{lcl}
\omega\left(0,0,1\right)\equiv\omega_{0} & = & \frac{1}{30}\left[10+7\left(e^{-\frac{\eta_{1}t}{2}}+e^{-\frac{\eta_{2}t}{2}}\right) + e^{-\frac{1}{2}\left(\eta_{1}+\eta_{2}\right)t}\right],\\
\omega\left(1,0,1\right)\equiv\omega_{1} & = & \frac{1}{60}\left[20-\left(e^{-\frac{\eta_{1}t}{2}}+e^{-\frac{\eta_{2}t}{2}}\right)- 13e^{-\frac{1}{2}\left(\eta_{1}+\eta_{2}\right)t}\right],\\
\omega\left(2,0,1\right)\equiv\omega_{2} & = & \frac{1}{60}\left[20-13\left(e^{-\frac{\eta_{1}t}{2}}+e^{-\frac{\eta_{2}t}{2}}\right)+ 11e^{-\frac{1}{2}\left(\eta_{1}+\eta_{2}\right)t}\right].
\end{array}\label{eq:tom-qutrit}
\end{equation}
It can be easily seen here that the tomogram obtained is normalized
as $\stackrel[m=0]{2}{\sum}\omega_{m}=1$ in Eq. (\ref{eq:tom-qutrit}).

\begin{figure}
\includegraphics[scale=0.6]{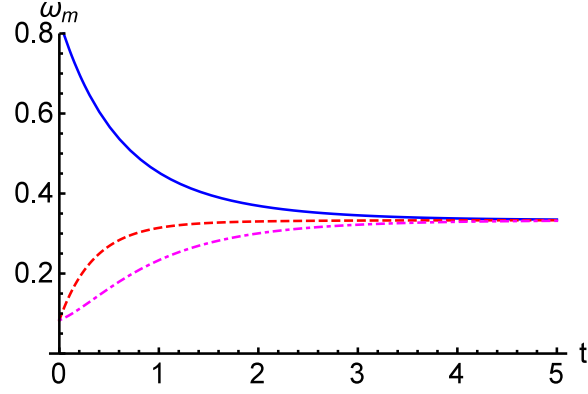}\protect\caption{\label{fig:qutrit}(Color online) The variation of different components
of tomogram for a qutrit state with time in the presence of a spontaneous
emission channel with Einstein coefficients $\eta_{1}=2$ and $\eta_{2}=4$.
The smooth (blue), dashed (red) and dot-dashed (magenta) lines correspond
to $\omega_{m}$ with $m=0,\,1,$ and 2, respectively.}

\end{figure}

For specific values of parameters of the spontaneous emission channel,
i.e., Einstein coefficients, the evolution of tomogram for the qutrit
state is shown in Fig. \ref{fig:qutrit}. The tomogram
shows that the noisy channel tends to randomize all the components
of the tomogram to one-third. We have already noted that once a tomogram is obtained for a finite dimensional system, it is possible to transform it to obtain the Wigner function for the system, and vice verse, but in an experiment we obtain tomograms. A lot of work has been devoted to the study of Wigner functions for finite dimensional systems
\cite{adam-wig1,adam-wig2,with-adam,with-jay}. Some efforts have also been made to study tomograms for finite dimensional coherent states \cite{with-adam,with-jay,manko-fss}. However, to the best of our knowledge, no such efforts had yet been made to study evolution of tomograms in noisy environment.

\section{Optical tomogram for a dissipative harmonic oscillator \label{sec:Optical-tomogram}}

In the end, we come to the tomogram of an infinite dimensional system, the harmonic oscillator. This is typical of a plethora 
of oscillatory and optical systems \cite{louis,perina-book}.   In Ref. \cite{comment13-ref1}, the quantum mechanics of the damped harmonic oscillator was examined,
from the perspective of a classical description of quantum mechanics \cite{tombesi96}. Use was made of the generating 
function method, resulting in the avoidance of the need to evaluate the Wigner function as an intermediary step for obtaining the tomogram. Further, in
\cite{comment13-ref2} the density matrix, state tomogram and Wigner function of a parametric oscillator were studied.   

Here, we construct the tomogram of the dissipative harmonic 
oscillator evolving under a Lindbladian evolution, in a phase sensitive reservoir \cite{op-tom-rho}. 
It would be pertinent to mention that tomographic reconstruction of Gaussian states evolving under
a Markovian evolution has also been considered in \cite{marzolino1}.
The dissipative harmonic oscillator can be described by the Hamiltonian 
\begin{equation}
H=H_{S}+H_{R}+H_{SR},\label{eq:hamil-opt-tom}
\end{equation}
where the system Hamiltonian $H_{s}$ of a harmonic oscillator is
described as 
\[
H_{S}=\frac{p^{2}}{2m}+\frac{1}{2}m\omega^{2}x^{2},
\]
while the reservoir Hamiltonian $H_{R}$ is given by 
\[
H_{R}=\sum_{j}\frac{p_{j}^{2}}{2m_{j}}+\frac{1}{2}m_{j}\omega_{j}^{2}x_{j}^{2},
\]
with the system-reservoir interaction Hamiltonian $H_{SR}$ as 
\[
H_{SR}=\sum_{j}c_{j}xx_{j}.
\]
Here, the reservoir is modeled as a bath of harmonic oscillators with
$c_{j}$ as the coupling constant. The dynamics of the system harmonic oscillator is obtained by tracing over the 
reservoir degrees of freedom. The optical tomogram from the Wigner
function can be obtained using \cite{op-tom} 
\begin{equation}
\begin{array}{l}
\omega\left(X,\theta\right)=\int W\left(X\cos\theta-p\sin\theta,X\sin\theta+p\cos\theta\right)dp,\end{array}\label{eq:tomogram_from_wigner}
\end{equation}
where $W\left(x,y,t\right)$ is the Wigner function. Similarly, the corresponding
Wigner function can also be reconstructed from the tomogram by inverse
Radon transformation.
The analytic expression of the tomogram for the system,
initially in the coherent state $\left|\beta\right\rangle,$ is 
\begin{equation}
\begin{array}{lcl}
\omega(X,\theta,t) & = & \sqrt{\frac{2}{\pi}}\frac{1}{\sqrt{\left(2NM+1\right)-\left(rMe^{-2i\theta}+{\rm c.c.}\right)}} \exp\left(-\frac{2\left(Re[\beta e^{i\theta}]e^{-kt}-X\right)^{2}}{\left(2NM+1\right)-\left(rMe^{-2i\theta}+{\rm c.c.}\right)}\right).
\end{array}\label{eq:optical_tomogram}
\end{equation}
Here, $k_{B}$ is the Boltzmann constant and $N=\frac{1}{\exp(\hbar \omega_{k}/k_{B}T) - 1}$ is the 
average thermal photon number of the environment at temperature $T$. 
Also, $r$ is the bath squeezing parameter, $Re[u]$ denotes the real part of $u$
and $M=1-\exp\left(-2kt\right),$ where $k$ is the dissipation coefficient, analogous to the spontaneous emission term. 

\begin{figure}
\includegraphics[angle=-90,scale=0.65]{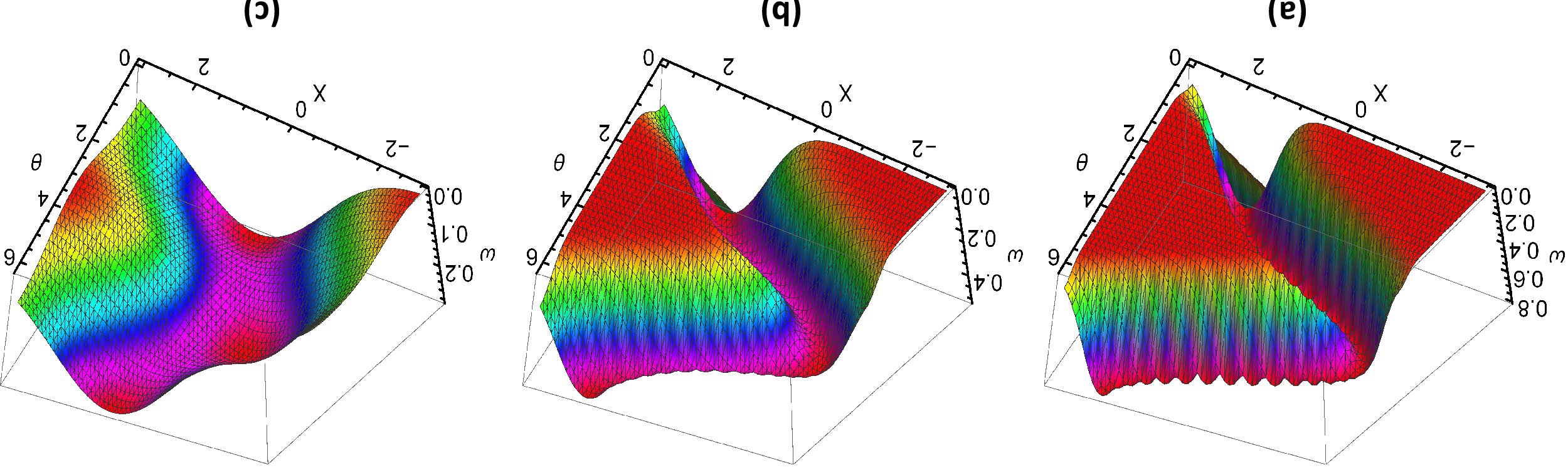}
\protect\caption{\label{fig:op-tom-3D}(Color online) The effect of interaction of the optical
tomogram with its environment is shown as a function of $X$ and $\theta$ for $N=5,\,r=1$ for the initial coherent state parameter $\beta=2$
at time $t=0,\,1,\,10$, respectively.}
\end{figure}

\begin{figure}
\includegraphics[scale=0.6]{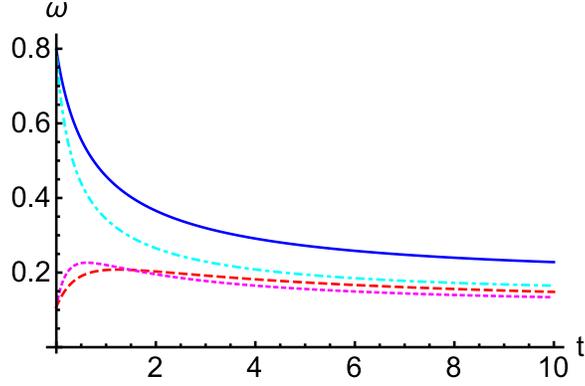}

\protect\caption{\label{fig:op-tom}(Color online) The tomogram of a dissipative harmonic oscillator varying with time is shown for the initial coherent state parameter $\beta=2$ and $\theta=\frac{\pi}{3}$.
The smooth (blue) and dashed (red) lines correspond to the tomogram
for $N=5,\,r=1$ for $X=1$ and 2, respectively. Similarly, the dot-dashed
(cyan) and dotted (magenta) lines correspond to the tomogram for $N=10,\,r=4$
for $X=1$ and 2, respectively.}
\end{figure}

In the corresponding figures for the tomogram with
specific values of different parameters, we can observe the decay
of the tomogram. Specifically, Fig. \ref{fig:op-tom-3D}-a shows the
tomogram of an initial coherent state, where we can see a beautiful valley
like shape surrounded by a mountain. Interestingly, a similar tomogram
has been observed for a binomial state of large dimension (cf. Fig. 2 in \cite{bin-tom}). However,
in Fig. \ref{fig:op-tom-3D}-b and c, we can see this sharp structure
gradually fade away due to interaction with its surrounding. 
  Thus, with increase in temperature, the effect of decoherence and dissipation,
due to the ambient environment, deteriorates the obtained tomogram.  
Further, Fig. \ref{fig:op-tom} illustrates the effect of change of average
thermal photon number and squeezing parameter, where in smooth (blue)
and dot-dashed (cyan) lines we can observe the enhancement of decay.
Similarly, dashed (red) and dotted (magenta) lines also show the effects
due to changes in bath parameters for another set of parameters.

\section{Conclusion \label{sec:Conclusion}}

Tomography is a powerful quantum state reconstruction tool. Its wide
applicability in obtaining quasidistribution functions, quantum process tomography and density matrix reconstruction
in quantum computation and communication is already established. 
However, these properties can get effected by the influence of the ambient environment.
Here, an effort has been made to study the evolution of tomograms
for different quantum systems, both finite and infinite dimensional,   under
general system-reservoir interactions,  
using the formalism of open quantum systems. The effect of the environment
on the finite dimensional quantum systems, both spin and number-phase states,
is to randomize the tomogram. For spin quantum states, single and two spin-$\frac{1}{2}$ states are considered with 
open quantum system effects. For the number-phase states, a general expression is obtained and is
illustrated through the example of a three level quantum (qutrit) system 
in a spontaneous emission channel. The increase in temperature tends to decohere 
the tomograms while squeezing is shown to be a  useful quantum resource. Besides this, a tomogram for a spin-1 pure 
quantum state is also obtained. Further, the tomogram for an infinite dimensional system, the
ubiquitous dissipative harmonic oscillator, is also studied.
The results obtained here are expected to have an impact on issues
related to quantum state reconstruction in quantum computation, communication and information processing.

  \section*{Acknowledgments}
Authors thank anonymous referee for constructive comments and for drawing their attention to a number of extremely relevant papers. 

\section*{Appendix 1}

The elements of the density matrix (\ref{eq:densitymatrix-vaccumbath}) are 
\begin{equation}
\begin{array}{lcl}
\rho_{ee}\left(t\right) & = & e^{-2\Gamma t}\rho_{ee}\left(0\right),\\
\rho_{ss}\left(t\right) & = & e^{-\left(\Gamma+\Gamma_{12}\right)t}\rho_{ss}\left(0\right) +  \frac{\left(\Gamma+\Gamma_{12}\right)}{\left(\Gamma-\Gamma_{12}\right)}\left(1-e^{-\left(\Gamma-\Gamma_{12}\right)t}\right)e^{-\left(\Gamma+\Gamma_{12}\right)t}\rho_{ee}\left(0\right),\\
\rho_{aa}\left(t\right) & = & e^{-\left(\Gamma-\Gamma_{12}\right)t}\rho_{aa}\left(0\right) +  \frac{\left(\Gamma-\Gamma_{12}\right)}{\left(\Gamma+\Gamma_{12}\right)}\left(1-e^{-\left(\Gamma+\Gamma_{12}\right)t}\right)e^{-\left(\Gamma-\Gamma_{12}\right)t}\rho_{ee}\left(0\right),\\
\rho_{gg}\left(t\right) & = & \rho_{gg}\left(0\right)+\left(1-e^{-\left(\Gamma+\Gamma_{12}\right)t}\right)\rho_{ss}\left(0\right) +  \left(1-e^{-\left(\Gamma-\Gamma_{12}\right)t}\right)\rho_{aa}\left(0\right)\\
 & + & \left[\frac{\left(\Gamma+\Gamma_{12}\right)}{2\Gamma}\left\{ 1-\frac{2}{\left(\Gamma-\Gamma_{12}\right)}e^{-\left(\Gamma+\Gamma_{12}\right)t}\right.
\left[\frac{\left(\Gamma+\Gamma_{12}\right)}{2}\left(1-e^{-\left(\Gamma-\Gamma_{12}\right)t}\right)+\frac{\left(\Gamma-\Gamma_{12}\right)}{2}\right]\right\} \\
 & + & \frac{\left(\Gamma-\Gamma_{12}\right)}{\left(\Gamma+\Gamma_{12}\right)}\left\{ \left(1-e^{-\left(\Gamma-\Gamma_{12}\right)t}\right) - \left.\frac{\left(\Gamma-\Gamma_{12}\right)}{2\Gamma}\left(1-e^{-2\Gamma t}\right)\right\} \right]\rho_{ee}\left(0\right),\\
\rho_{es}\left(t\right) & = & e^{-i\left(\omega_{0}-\Omega_{12}\right)t}e^{-\frac{1}{2}\left(3\Gamma+\Gamma_{12}\right)t}\rho_{es}\left(0\right),\\
\rho_{ea}\left(t\right) & = & e^{-i\left(\omega_{0}+\Omega_{12}\right)t}e^{-\frac{1}{2}\left(3\Gamma-\Gamma_{12}\right)t}\rho_{ea}\left(0\right),\\
\rho_{eg}\left(t\right) & = & e^{-2i\omega_{0}t}e^{-\Gamma t}\rho_{eg}\left(0\right),\\
\rho_{sa}\left(t\right) & = & e^{-2i\Omega_{12}t}e^{-\Gamma t}\rho_{sa}\left(0\right),\\
\rho_{sg}\left(t\right) & = & e^{-i\left(\omega_{0}+\Omega_{12}\right)t}e^{-\frac{1}{2}\left(\Gamma+\Gamma_{12}\right)t} \left[\rho_{sg}\left(0\right)+\frac{\left(\Gamma+\Gamma_{12}\right)}{\left(\Gamma^{2}+4\Omega_{12}^{2}\right)}\left(\left\{ 2\Omega_{12}e^{-\Gamma t}\sin\left(2\Omega_{12}t\right)\right.\right.\right.\\
 & + & \Gamma\left.\left(1-e^{-\Gamma t}\cos\left(2\Omega_{12}t\right)\right)\right\} + i\left\{ 2\Omega_{12}\left(1-e^{-\Gamma t}\cos\left(2\Omega_{12}t\right)\right) - \left.\Gamma e^{-\Gamma t}\sin\left(2\Omega_{12}t\right)\right\}\rho_{es}\left(0\right)\right],\\
\rho_{ag}\left(t\right) & = & e^{-i\left(\omega_{0}-\Omega_{12}\right)t}e^{-\frac{1}{2}\left(\Gamma-\Gamma_{12}\right)t}\left[\rho_{ag}\left(0\right)-\frac{\left(\Gamma-\Gamma_{12}\right)}{\left(\Gamma^{2}+4\Omega_{12}^{2}\right)}\right. \left(\left\{ 2\Omega_{12}e^{-\Gamma t}\sin\left(2\Omega_{12}t\right)\right.\right.\\
 & + & \left.\Gamma\left(1-e^{-\Gamma t}\cos\left(2\Omega_{12}t\right)\right)\right\} - i\left\{ 2\Omega_{12}\left(1-e^{-\Gamma t}\cos\left(2\Omega_{12}t\right)\right) - \left.\left.\Gamma e^{-\Gamma t}\sin\left(2\Omega_{12}t\right)\right\} \right)\rho_{ea}\left(0\right)\right].
\end{array}\label{eq:matrix-elements}
\end{equation}
Here, all the matrix elements are written in the dressed state basis, which is connected with the bare state basis by
\[
\begin{array}{lcl}
\left|g\right\rangle  & = & \left|g_{1}\right\rangle \left|g_{2}\right\rangle ,\\
\left|s\right\rangle  & = & \frac{1}{\sqrt{2}}\left(\left|e_{1}\right\rangle \left|g_{2}\right\rangle +\left|g_{1}\right\rangle \left|e_{2}\right\rangle \right),\\
\left|a\right\rangle  & = & \frac{1}{\sqrt{2}}\left(\left|e_{1}\right\rangle \left|g_{2}\right\rangle -\left|g_{1}\right\rangle \left|e_{2}\right\rangle \right),\\
\left|e\right\rangle  & = & \left|e_{1}\right\rangle \left|e_{2}\right\rangle .
\end{array}
\]
Further,
\[
\begin{array}{lcl}
\Omega_{ij} & = & \frac{3}{4}\sqrt{\Gamma_{i}\Gamma_{j}}\left[-\left[1-\left(\hat{\mu}\cdot\hat{r}_{ij}\right)^{2}\right]\frac{\cos\left(k_{0}r_{ij}\right)}{k_{0}r_{ij}} + \left[1-3\left(\hat{\mu}\cdot\hat{r}_{ij}\right)^{2}\right]\left(\frac{\sin\left(k_{0}r_{ij}\right)}{\left(k_{0}r_{ij}\right)^{2}}+\frac{\cos\left(k_{0}r_{ij}\right)}{\left(k_{0}r_{ij}\right)^{3}}\right)\right],
\end{array}
\]
where $\hat{\mu}=\hat{\mu}_{1}=\hat{\mu}_{2}$ are the unit vectors
along the atomic transition dipole moments, $\hat{r}_{ij}=\hat{r}_{i}-\hat{r}_{j}$,
and $k_{0}=\frac{\omega_{0}}{c}$ with $\omega_{0}=\frac{\omega_{1}+\omega_{2}}{2}$;
the spontaneous emission rate is 
\[
\begin{array}{lcl}
\Gamma_{i} & = & \frac{\omega_{i}^{3}\mu_{i}^{2}}{3\pi\epsilon\hbar c^{3}},\end{array}
\]
and the collective incoherent effect due to the dissipative multi-qubit
interaction with the bath is 
\[
\begin{array}{lcl}
\Gamma_{ij} & = & \Gamma_{ji}=\sqrt{\Gamma_{i}\Gamma_{j}}F\left(k_{0}r_{ij}\right),\end{array}
\]
for $i\neq j$ with 
\[
\begin{array}{lcl}
F\left(k_{0}r_{ij}\right) & = & \frac{3}{2}\left[\left[1-\left(\hat{\mu}\cdot\hat{r}_{ij}\right)^{2}\right]\frac{\sin\left(k_{0}r_{ij}\right)}{k_{0}r_{ij}} + \left[1-3\left(\hat{\mu}\cdot\hat{r}_{ij}\right)^{2}\right]\left(\frac{\cos\left(k_{0}r_{ij}\right)}{\left(k_{0}r_{ij}\right)^{2}}-\frac{\sin\left(k_{0}r_{ij}\right)}{\left(k_{0}r_{ij}\right)^{3}}\right)\right].
\end{array}
\]
Further, for the case of identical qubits, as  considered here,
$\Omega_{12}=\Omega_{21}$, $\Gamma_{12}=\Gamma_{21}$, and $\Gamma_{1}=\Gamma_{2}=\Gamma$.

\end{document}